\documentclass[preprint]{aastex}
\usepackage{amsmath, amsfonts, amssymb}
\usepackage{graphicx}
\usepackage{subfigure}
\usepackage{multirow}
\usepackage{hyperref}
\usepackage{natbib}    
\begin{document}

\title{The Allen Telescope Array Search for Electrostatic Discharges on Mars}

\author{Marin M. Anderson\altaffilmark{1}, Andrew P.V. Siemion\altaffilmark{1}, William C. Barott\altaffilmark{2}, Geoffrey C. Bower\altaffilmark{1}, Gregory T. Delory\altaffilmark{3}, Imke de Pater\altaffilmark{1}, and Dan Werthimer\altaffilmark{3}}
\altaffiltext{1}{Department of Astronomy and Radio Astronomy Laboratory, University of California, Berkeley; \href{mailto:marinanderson@berkeley.edu}{marinanderson@berkeley.edu}}
\altaffiltext{2}{Embry-Riddle Aeronautical University, Daytona Beach, Florida}
\altaffiltext{3}{Space Sciences Laboratory, University of California, Berkeley}

\begin{abstract}

The Allen Telescope Array was used to monitor Mars between 9 March and 2 June 2010, over a total of approximately 30 hours, for radio emission indicative of electrostatic discharge.  The search was motivated by the report from \citet{Ruf:2009uy} of the detection of non-thermal microwave radiation from Mars characterized by peaks in the power spectrum of the kurtosis, or kurtstrum, at 10 Hz, coinciding with a large dust storm event on 8 June 2006.  For these observations, we developed a wideband signal processor at the Center for Astronomy Signal Processing and Electronics Research (CASPER).  This 1024-channel spectrometer calculates the accumulated power and power-squared, from which the spectral kurtosis is calculated post-observation.  Variations in the kurtosis are indicative of non-Gaussianity in the signal, which can be used to detect variable cosmic signals as well as radio frequency interference (RFI).  During the three month period of observations, dust activity occurred on Mars in the form of small-scale dust storms, however no signals indicating lightning discharge were detected.  Frequent signals in the kurtstrum that contain spectral peaks with an approximate 10 Hz fundamental were seen at both 3.2 and 8.0 GHz, but were the result of narrowband RFI with harmonics spread over a broad frequency range.
\end{abstract}


\section{Introduction} \label{introduction} 
Evidence of electrostatic discharge in the Martian atmosphere, in the form of fluctuations in non-thermal microwave radiation, visible as spectral peaks in the kurtosis, was reported by \citet{Ruf:2009uy}.  During the period of observations of Mars between 22 May to 16 June 2006, the periodic fluctuations in non-thermal emission were detected once during a single event, and were attributed to a massive, 35-kilometer deep dust storm on 8 June 2006.  Large-scale dust storms and smaller-scale dust devils are prevalent on Mars, and the resulting large-scale charge distributions in the Martian atmosphere due to frictional charging of dust grains are theorized to reach strengths large enough to cause electrostatic discharge \citep{Eden:1973gj, Farrell:2004ws, Farrell:1999ud}.  \citet{Gurnett:2010} conducted a similiar search between 4.0 and 5.5 MHz for impulsive radio signals from lightning discharges in Martian dust storms over a period of five years using the radar receiver on Mars Express.  Two large dust storms, including the 8 June 2006 storm, and many smaller storms occurred during the search period, however no signals indicating lightning were detected.  The search by \citet{Ruf:2009uy} for lightning discharge was done using DSS-13, a 34-meter parabolic antenna that is part of the NASA Deep Space Network (DSN), operating in the band 8470 to 8490 MHz.  The low-frequency spectral peaks at approximately 10 Hz and multiples of 10 Hz in the kurtosis seen during the large dust storm event were explained as modes of Mars' Schumann resonances, induced by electrostatic discharge.  Schumann resonances are seen on Earth as extremely low-frequency emission, caused by terrestrial lightning, which propagate in the cavity formed by the Earth's surface-ionosphere boundaries \citep{Yang:2006wu}; the proposed
mechanism for producing Schumann resonances is the same on Mars.  These resonances were first predicted and solved analytically by \citet{Schumann:1952}.  The possible detection of lightning discharges on Mars by \citet{Ruf:2009uy} motivated the observations of Mars reported here.



In this paper, 
we discuss the potential for the kurtosis as a means of identifying and characterizing astrophysical emissions, in addition to excising RFI (\S \ref{kurtosis}); 
we describe the special wideband signal processor developed specifically for our observations of Mars, with the Allen Telescope Array (\S \ref{observations} and \ref{instrumentation}); 
and we detail our results, specifically, the recurring, periodic signal in the spectrum of the kurtosis caused by narrowband RFI which appears over multiple frequency bands
(\S \ref{analysis} and \ref{discussion}).  We summarize in \S \ref{conclusion}.

Our observations were made as part of a joint campaign with a DSN antenna
\citep{2010DPS....42.3013K}.  No convincing evidence for electrostatic discharge from Mars
was found in either search.



\section{Spectral Kurtosis Estimator \label{kurtosis}}
The kurtosis is the ratio between the fourth moment and the square of the second moment.  Its insensitivity to Gaussian-distributed noise (i.e. thermal noise) 
provides a means of distinguishing non-Gaussian and Gaussian signals.  In general, radio
astronomy use of the kurtosis has been largely limited to radio frequency interference excision; as a measure of non-Gaussianity, kurtosis algorithms are effective at identifying and removing terrestrial RFI contaminating spectra of astrophysical sources \citep{Nita:2007cd}.  The use of the kurtosis as a means for detecting the non-thermal emission on Mars associated with electrostatic discharge presents an alternate use of kurtosis 
that might be applicable to other sources such as pulsar giant pulses and other short-duration
astrophysical phenomena.  
The spectral kurtosis (SK) estimator and its statistical variance, detailed in \citet{Nita:2007cd} and used in our experiment, define a threshold for identifying RFI or non-Gaussian astrophysical signals from data.

The algorithm for computing the SK estimator, taken from \citet{Nita:2007cd}, uses the ratio of the accumulated power-squared, $S_{2}$, and power, $S_{1}$, defined as:

\begin{equation}
S_{1} = \sum^{M}_{i=1} \hat{P}_{ki}, \indent S_{2} = \sum^{M}_{i=1} \left( \hat{P}_{ki} \right)^{2},
\end{equation}

\noindent where $\hat{P}_{k}$ is the power spectral density estimate; the sum over $M$ is the number of accumulations for $N$ time-domain samples (subscript $k$ runs from 0, ..., $N/2$).  The SK estimator, $\hat{V_{k}^{2}}$, is given by\footnote{\citet{Nita:2010gy} adjusts the SK estimator defined previously in \citet{Nita:2007cd}, rescaling it by $(M+1)/M$.  In the case of our data, this rescaling would have a negligible effect.}

\begin{equation} \label{skestimator}
\hat{V}_{k}^{2} = \frac{M}{M-1} \left(M\frac{1}{f} \frac{S_{2}}{(S_{1})^{2}} - 1 \right).
\end{equation}

\noindent Here we have included the scaling coefficient $1/f$ to normalize the ratio of the accumulated power-squared and power; this normalization is necessary because of the design of our instrument
.  For our observations, $f = 144$ and $M = 128$.  In the presence of a Gaussian signal, the SK estimator takes on a value of 1; a departure in the signal from Gaussianity moves the SK estimator away from unity.  We define the $3\sigma$ bounds around 1 as the detection threshold past which the spectral kurtosis indicates a significant non-Gaussian signal.  The variance of Eq. \ref{skestimator} is 

\begin{equation}
\mbox{Var}(\hat{V}_{k}^{2}) \simeq \left\{ \begin{array}{rl}
24/M, &\mbox{ $k = 0 \mbox{, } N/2,$} \\
4/M, &\mbox{ $k = 1, \mbox{ ..., } (N/2 - 1),$}
\end{array} \right.
\end{equation}

\noindent thus the $3\sigma$ bounds are given by $1 \pm 0.53$ (in all but the DC and Nyquist frequency bins).


\subsection{The Kurtstrum}
We define the kurtstrum $K$ as

\begin{equation} \label{kurtstrumgen}
K = \left| \mathcal{F} \Biggl\{ \hat{V}^{2}_{k} \Biggr\} \right|^{2}
\end{equation}

\noindent which is computed by taking the power spectrum of the spectral kurtosis over a single frequency channel $k$, where $\mathcal{F} \{ \}$ is defined as the Fourier transform.  The kurtstrum, or power spectrum of the kurtosis, identifies low-frequency modulations in the spectral kurtosis as a function of time. 
Due to the high number of channels measured by the spectrometer, from our data we calculate the mean-subtracted kurtstrum for four individual bands, each band consisting of a quarter of all the $N = 1024$ frequency channels:

\begin{equation} \label{kurtstrum}
K_{\mu} = \left| \mathcal{F} \Biggl\{ \sum_{k=i}^{j}{\hat{V}^{2}_{k}} - \frac{1}{j} \sum_{k=i}^{j}{\hat{V}^{2}_{k}}   \Biggr\} \right|^{2}, \indent \begin{cases}
k = 0, \mbox{ ..., } N/4, \\
k = (N/4 + 1), \mbox{ ..., } N/2, \\
k = (N/2 + 1), \mbox{ ..., } 3N/4, \\
k = (3N/4 + 1), \mbox{ ..., } N.
\end{cases}
\end{equation} 

\noindent 
As shown in Eq. \ref{kurtstrum}, the mean-subtracted kurtstrum is computed by collapsing the $N = 1024$ frequency channels of the spectral kurtosis (as a function of frequency and time) into four bands; summing over the frequency channels and subtracting off the mean for each band; and computing the power spectrum for each one second of data (giving a frequency resolution of 1 Hz).  The low-frequency modulation of the kurtosis which appears over minute-long timescales in our Mars observations (see Section \ref{analysis}) was also seen in the \citet{Ruf:2009uy} observations. In \citet{Ruf:2009uy} the low-frequency modulation coincided with a large-scale Martian dust storm and showed spectral peaks near the frequencies of the fundamental modes of Mars' Schumann resonance.

\section{Observations} \label{observations}
Observations of Mars were done with the Allen Telescope Array (ATA) at the Hat Creek Observatory in California, an array of 42 6.1 meter-equivalent offset Gregorian antennas \citep{Welch:2009cq}.  Only a subset of these antennas was used for this experiment; Table \ref{senstable} lists the number used in each observation.  Between 9 March and 2 June 2010, Mars was observed on 7 days for roughly 4.25 hours per day, for a total of approximately 30 hours of observation.  Additional observations were done on the pulsar PSR B0329+54 and OH/H$_{2}$O masers. 
Initial testing of the spectrometer with the ATA took place on 10 February 2010, on a number of sources, including Jupiter (both on- and off-source), PSR B0329+54, PSR B0531+21, and a few terrestrial sources, such as the RADARSAT-1 satellite.  
The Mars observation dates were chosen to coincide with JPL observations using DSS-13 at 8.0 GHz to similarly search for electrostatic discharge on Mars \citep{2010DPS....42.3013K}.  Both searches were conducted during the off-season for Martian dust storms, which reach their peak during Mars' southern spring and summer.  However, small-scale convective dust activity, such as dust devils, in contrast to the seasonally dependent large-scale planet-wide dust storms, are active during most of the year \citep{Cantor:2007ur}.

Observation dates and times are given in Table \ref{obstable}.  
Simultaneous observations were done on Mars with two beamformers (BF1, BF2), tuned to 3.2 GHz and 8.0 GHz; no off-source data were taken while observing Mars.  
Our intention was that overlap with the JPL DSN observations would provide rejection
of local RFI; however, in practice the amount of overlap in observations was limited to less than five hours.
Two observing frequencies were used in order to determine whether any variability
was truly broadband and reject narrowband RFI that could corrupt the results.
The synthesized beam of the ATA was $\sim 100$\arcsec at 3.2 GHz and $\sim 40$\arcsec
at 8.0 GHz; both beam sizes are significantly larger than the angular diameter of
Mars at the time of the observations, which ranged from 5 to 12\arcsec.

Table \ref{senstable} gives the effective system equivalent flux density (SEFD) for each observation.  This was calculated using the equal noise power combining (ENPC) method, where the antennas are weighted by equal thermal noise power.  This is in contrast to the theoretically-best maximal ratio combining (MRC) method, in which the antennas are weighted according to their signal-to-noise ratios.  In this case, ENPC is sufficient, as it yields values within $\sim 6$\% of the best possible sensitivity values given by MRC.  Phase and amplitude calibration errors are expected to degrade performance by no more than a few percent.  Typical sensitivities in a 1.25 ms integration over the full band of the spectrometer were 1.6 Jy at 3.2 GHz and 2.2 Jy at 8.0 GHz.


\section{The High Time Resolution Kurtosis Spectrometer \label{instrumentation}}

The Allen Telescope Array (ATA) is equipped with three time-domain, dual-polarization beamformers \citep{Barott:2011}, two of which were available for use in this experiment.  These instruments delay the signal path from each antenna in the array such that the signals from all antennas can be summed coherently to form a single-pixel synthesized beam with a collecting area given by the sum of the individual antenna collecting  areas and a beamwidth given by the baseline separation of array elements.  The ATA beamformers digitize and process a 104.8 MHz total bandwidth, of which $\sim$72 MHz is relatively unaffected by filter roll-off and phase and amplitude match errors.  The degradation is slow, however, and in most cases the entire band is used.
The synthesized beam for each beamformer is independently steerable across the primary beam and tunable across the entire 500 MHz to 10 GHz receiver band.  The ATA beamformers output a digital stream of complex baseband voltage data, delivered as 8 bits real / 8 bits imaginary in dual polarization at 104.8 Msamples/sec, and can be sent over serial or packetized links on a variety of media.  

\begin{figure}[htb]
\includegraphics[width=1.0\textwidth]{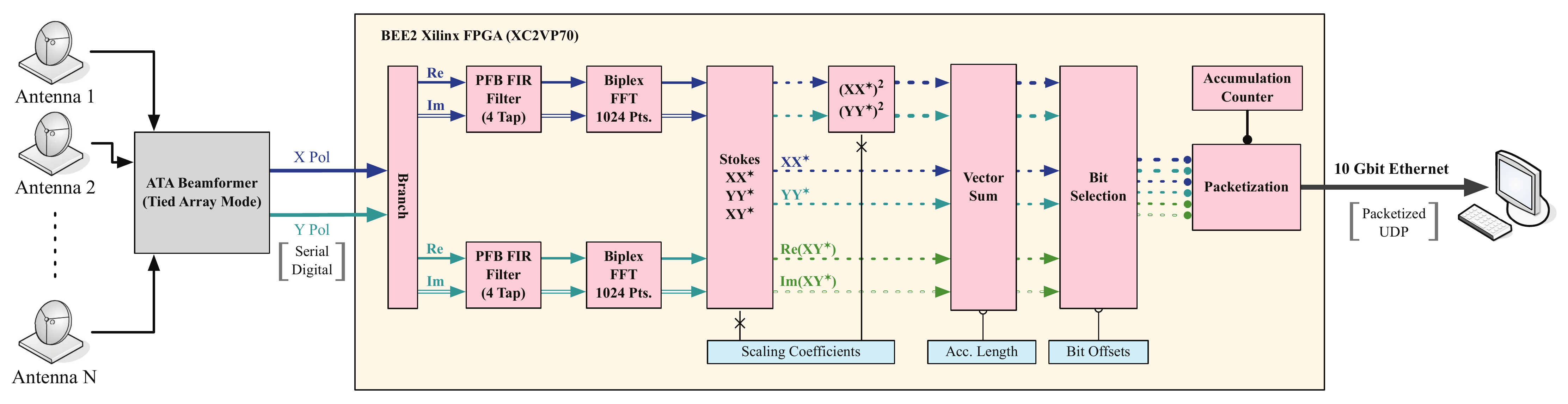}
\caption{
The ATA beamformer feeds a digital stream of complex baseband data to HiTREKS.  After channelization and accumulation, spectra are transmitted to a PC over 10 GbE.
\label{fig:hitreks}
}
\end{figure}

For this experiment, we designed a fast-dump spectrometer to reside on a spare field programmable gate array chip (FPGA) available on each beamformer, receiving the digital synthesized beam directly over on-board inter-chip communication.  The digital design for this instrument, hereafter the high time resolution kurtosis spectrometer (HiTREKS), was developed using the Center for Astronomy Signal Processing and Electronics Research (CASPER) \citep{Werthimer:2011ux} open source signal processing libraries and toolflow.  A block diagram of HiTREKS is shown in Figure~\ref{fig:hitreks}.  A serial digital stream of complex baseband voltage data is fed over interchip links to the HiTREKS FPGA, where a 1024 point 4-tap polyphase filterbank channelizes the input stream.  Following channelization, power $\hat{P}_{ki}=\{XX^*,YY^*\}$,  power squared $(\hat{P}_{ki})^2=\{(XX^*)^2,(YY^*)^2\}$ and cross product terms $XY^*$ are computed, accumulated and output over a 10 GbE link as UDP packets sent to a host computer.  Scaling coefficient and bit offset inputs provide control over the data stream magnitude as it propagates through the digital logic, and the accumulation length setting controls integration time.  Time keeping is accomplished using an accumulation counter, slaved to the beamformer's absolute time standard and prepended to each UDP packet.  Our development effort leveraged two previously designed CASPER instruments, the Berkeley ATA Pulsar Processor \citep{McMahon:2008tk} and a wideband kurtosis spectrometer built for solar observations \citep{Gary:2010wh}.

For this experiment, HiTREKS was configured for 1.25 millisecond integrations, with a resultant data rate of $\sim$7 MBps per synthesized beam, including metadata overhead.  In total, these observations produced about 1.5 TB.  All of these data are archived in pseudo-FITS format and are available for analysis by request to the authors.



\subsection{Diagnostic Observations}
\begin{figure}[htb]
\includegraphics[width=1.0\textwidth]{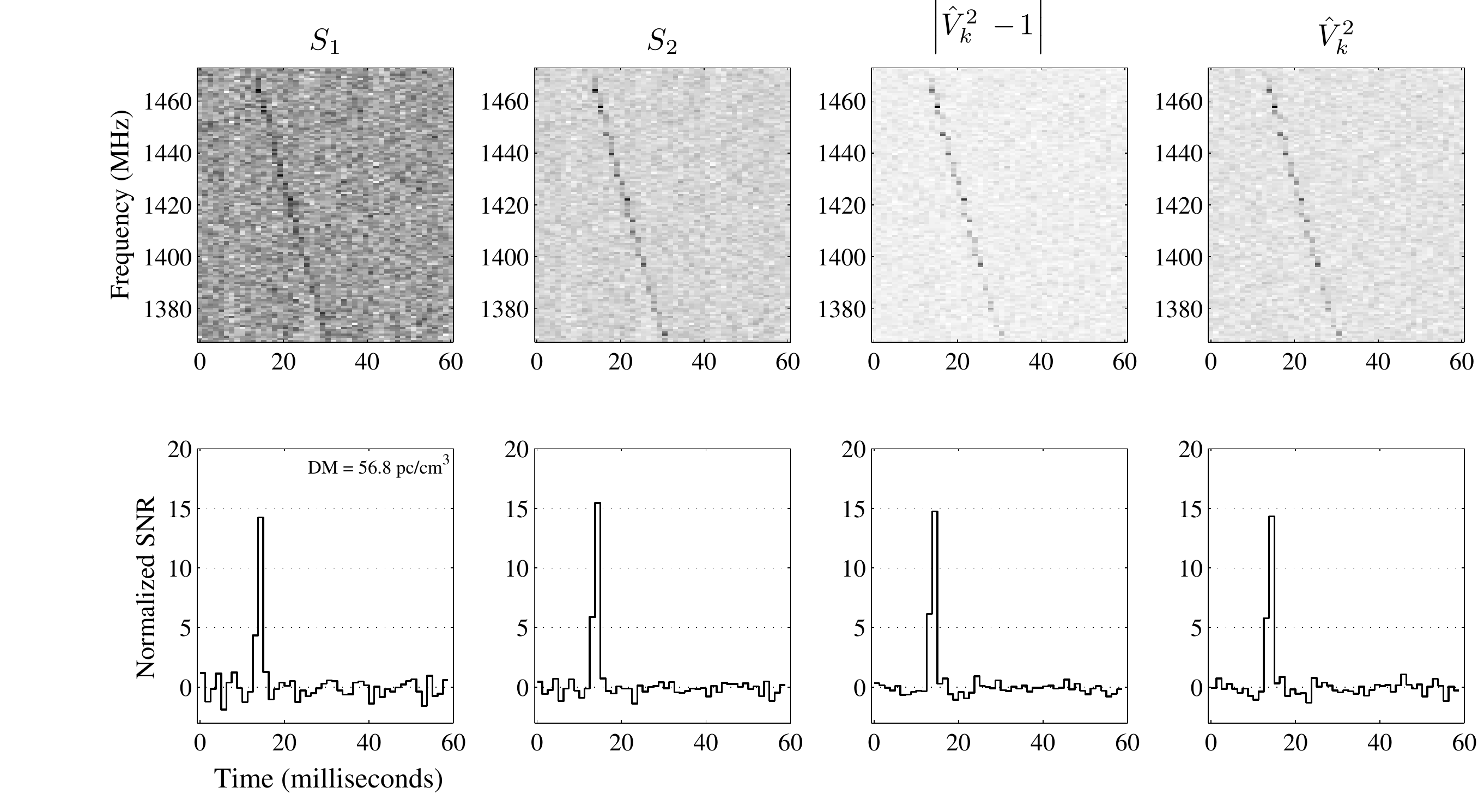}
\caption{
A single giant pulse from the Crab pulsar as observed with HiTREKS.  Shown are detections using four quantities derived from the HiTREKS spectral products, each of which is defined in Section \ref{kurtosis}.  Top panels show the dispersed pulse across the observing band, bottom panels show incoherently dedispersed pulse profiles.  These data were taken at a center frequency of 1420 MHz with the nominal HiTREKS configuration (Table \ref{tab:hitreks_specs}) and both linear polarizations have been summed.  The pulse shown was the brightest detected in a 30 minute observation. 
\label{fig:crabgps}
}
\end{figure}

\begin{figure}[htb]
\includegraphics[width=1.0\textwidth]{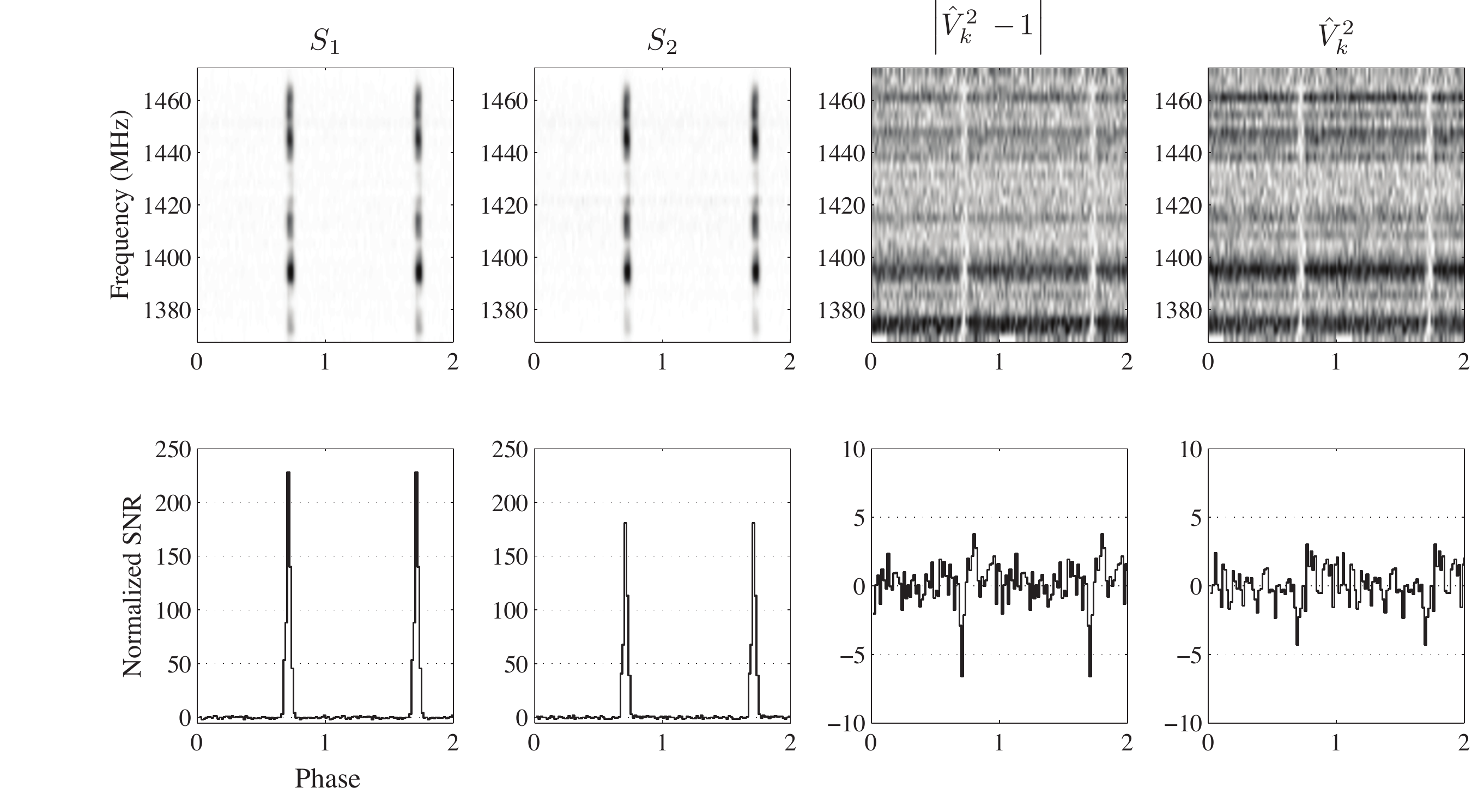}
\caption{
Folded pulse profiles for a 9.6 minute observation of the pulsar B0329$+$54.  Shown are detections using four quantities derived from the HiTREKS spectral products, each of which is defined in Section \ref{kurtosis}. Top panels show folded pulse profiles across the observed band. Bottom panels show the incoherently dedispersed pulse profile summed over the observed band.  These data were taken at a center frequency of 1420 MHz with the nominal HiTREKS configuration (Table \ref{tab:hitreks_specs}) and both linear polarizations have been summed. No interference excision has been performed.  Two full turns are plotted for clarity.
\label{fig:0329profs}
}
\end{figure}

\begin{figure}[h!]
    \begin{center}
      \subfigure [] {\label{radarsatpower} \includegraphics[width=.9\textwidth]
  {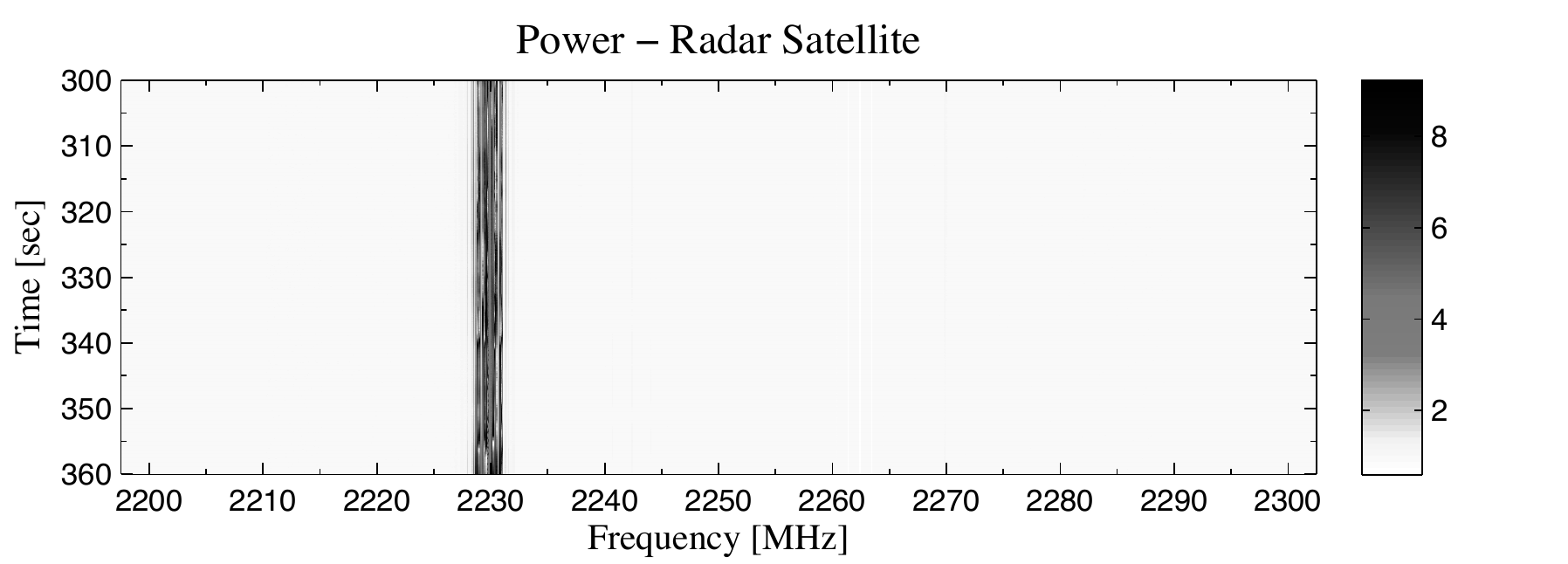}}
      \subfigure [] {\label{radarsatkurt} \includegraphics[width=.9\textwidth]
  {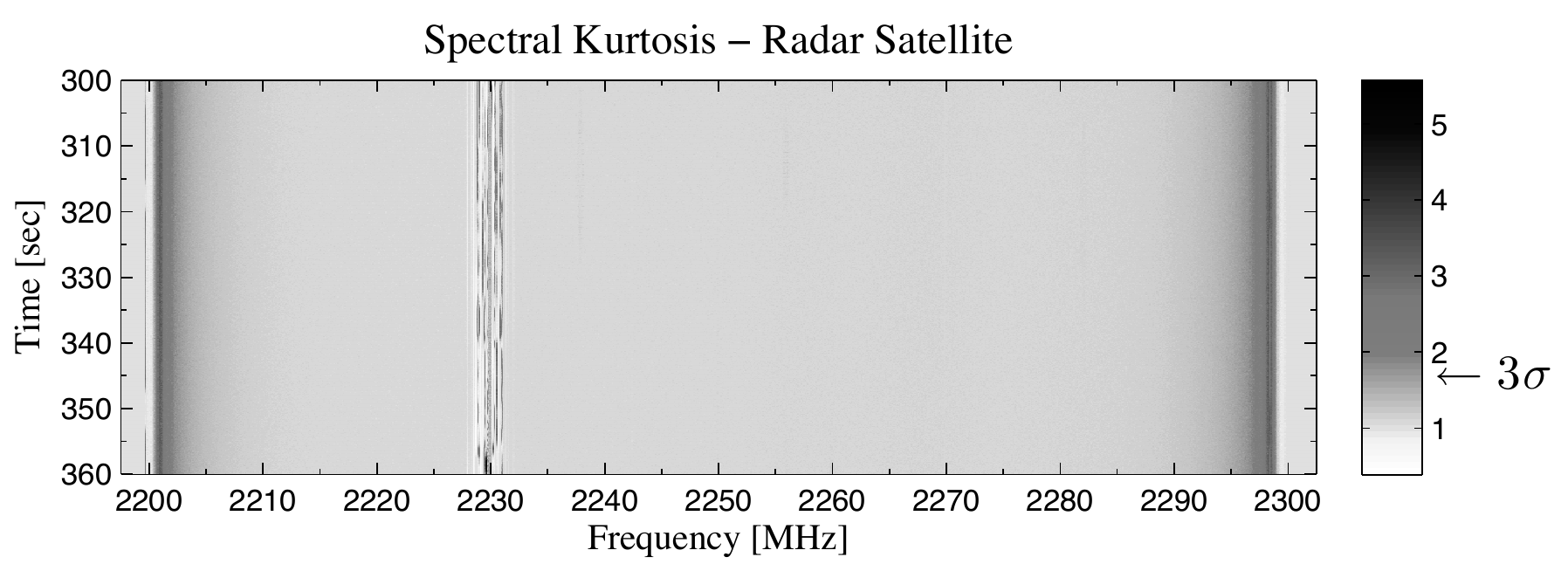}}
    \end{center}
  \caption{Accumulated power (top) and spectral kurtosis (bottom) as a function of sky frequency and time while tracking the RADARSAT-1 2230 MHz communication downlink transponder. \label{radarsat}}
\end{figure}

Prior to our Mars observing campaign, HiTREKS was used to observe a number of diagnostic sources to test its operability.  These sources included both man-made producers of non-thermal emission, satellite communication transponders and terrestrial radar installations, as well as astrophysical sources of highly time variable emission, several bright pulsars and masers.  Figures \ref{fig:crabgps}, \ref{fig:0329profs} and \ref{radarsat} show illustrative results from these observations.  

The Crab pulsar is known to intermittently produce individual pulses that are orders of magnitude brighter than its average emission \citep{Hankins:2003}, and it is frequently used as a diagnostic source for high time resolution radio astronomy instruments.  Figure \ref{fig:crabgps} shows a single dispersed pulse from the Crab pulsar as detected by four quantities derived from the HiTREKS spectral products, each of which is defined in Section \ref{kurtosis}.  These data were taken at a center frequency of 1420 MHz, and both linear polarizations have been summed.  The pulse shown was the brightest detected in a 30 minute observation.  In the upper panels, the pulse appears chirped due to the frequency dependent index of refraction of the interstellar medium (ISM).  The lower panels show ``dedispersed'' profiles of this pulse, in which the dispersive effects of the ISM have been corrected for, based on the known dispersion measure (DM) of the Crab pulsar, $~56.8$ pc/cm$^3$, by sliding frequency channels relative to one another and summing across the band.  Giant pulses from the Crab pulsar have been studied extensively at these frequencies e.g. \citet{Karuppusamy:2010et}, and the brightest of these pulses have observed pulse widths 1-2 orders of magnitude shorter than our integration time and exhibit significant structure at time scales greater than the Crab pulsar scattering time, $\tau_{\rm{s - 1400 MHz}}\sim 1{ }\mu $s.  The strong detection in the kurtosis excess and kurtosis (right two panels of Figure \ref{fig:crabgps}) illustrate the strong non-Gaussianity we expect in our 1.25 ms integrations based on these known properties of Crab giant pulses.

Figure \ref{fig:0329profs} shows pulse profiles of the bright pulsar B0329+54 folded at the known period of $\sim$0.71s.  These observations were also taken with a center frequency of 1420 MHz, and we show the same four quantities as in Figure \ref{fig:crabgps}, also with summed polarizations.  In contrast with the Crab pulsar results, we see in Figure \ref{fig:0329profs} very strong detections in the $S_1$ and $S_2$ moments, but little or no detection in the kurtosis related quantities.  B0329+54 is not known to produce giant pulses, and its periodic emission can be reasonably described by the amplitude modulated noise (AMN) model of \citet{Rickett:1975}.  We can thus consider B0329$+$54 pulsed emission to be Gaussian noise modulated by a Gaussian envelope, of width approximately equal to the duty cycle of the pulsar $W_{10} \sim 30$ ms \citep{Lorimer:1995tr}.  In this case the Gaussian pulse envelope is well sampled by our 1.25 ms integrations, as evidenced by the only very weak detection of pulsed emission in the two kurtosis-related quantities (right two plots of Figure \ref{fig:0329profs}).  Variations in intensity across the band are due to interstellar scintillation.

Figure \ref{radarsat} shows the canonical property of spectral kurtosis, the ability to identify the non-Gaussianity associated with man-made radio interference.  Here we observed the 2230 MHz transponder of the RADARSAT-1 satellite \citep{ieee:radarsat} using a center frequency tuning of 2250 MHz.  The upper panel shows total power ($S_1$) as a function of frequency and time, the bottom panel shows the spectral kurtosis ($\hat{V}_{k}^{2}$) over the same frequency/time plane.  As expected, the spectral kurtosis values in the transponder band indicate strong non-Gaussianity.



\begin{figure}[h!]
\begin{center}
\subfigure[]{\label{psk1} \includegraphics[width=1\textwidth]
  {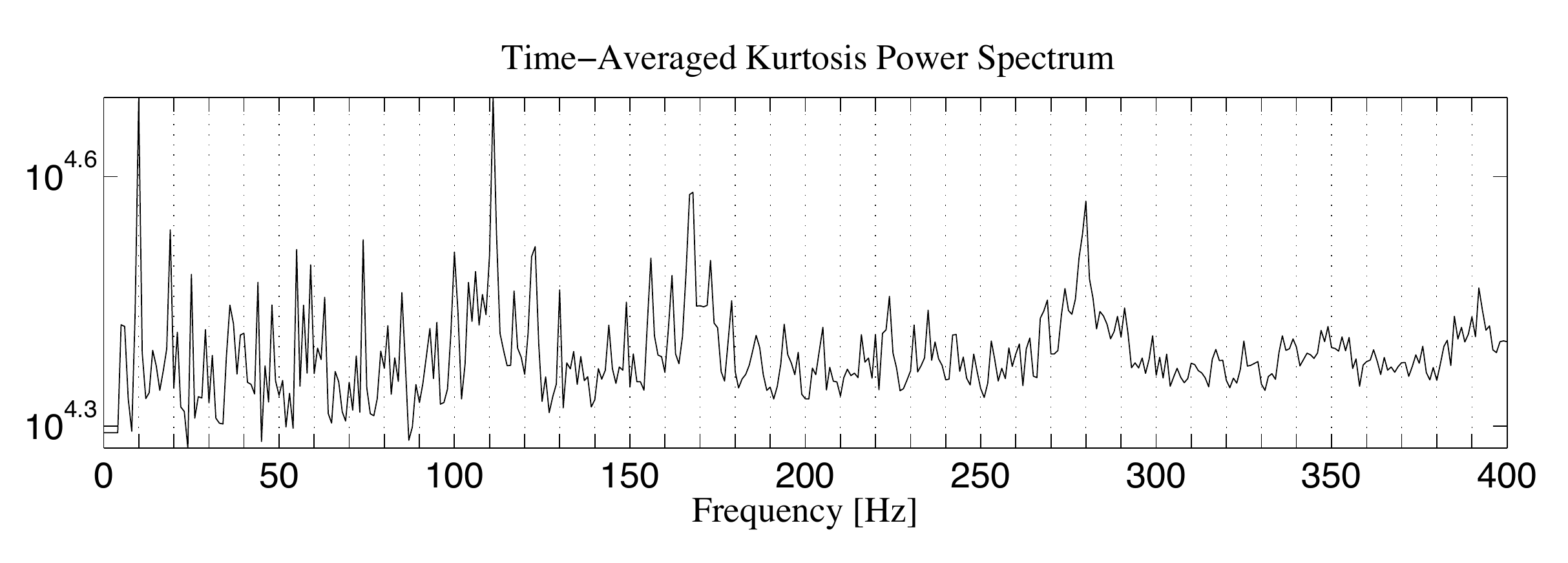}}
\subfigure[]{\label{psp1} \includegraphics[width=1.01\textwidth]
  {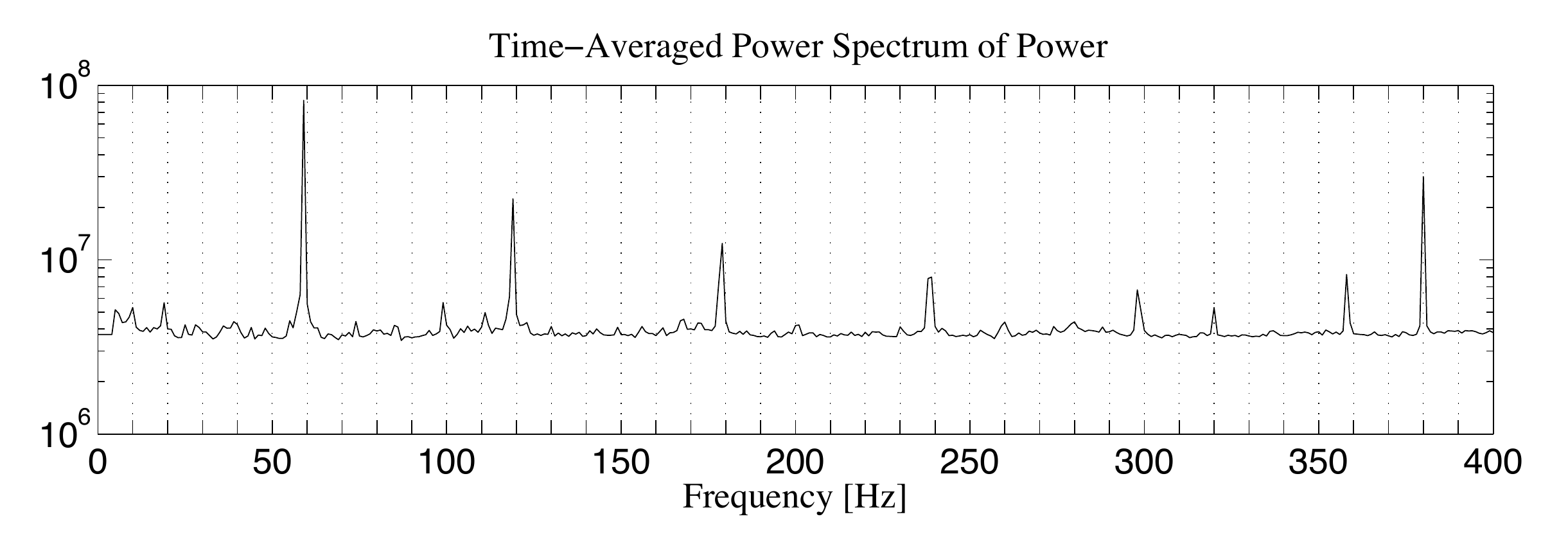}}
\end{center}
\caption{Kurtstrum over an approximately two hour-long period of Mars observations (top) and equivalent plot for the power (bottom).  The prominent 60 Hz features in the time-averaged power plot (due to US mains power) are almost completely gone in the kurtstrum. \label{ps1}}
\end{figure}

Figure \ref{ps1} shows the power spectrum of the kurtosis (top) and of the power (bottom), averaged over a two-hour period of data from observations of Mars on 23 March 2010.  The kurtstrum contains significant spectral peaks at 10 Hz, 110 Hz, 170 Hz, etc., but does not exhibit the 60 Hz and higher harmonics noise which is dominant in the power.  The kurtosis is insensitive to the 60 Hz modulation that results from the frequency of electric power distribution in the United States.  The 10 Hz and higher harmonics features that are dominant in the kurtstrum are a result of RFI.


\section{Analysis} \label{analysis}
Post-observation, the spectral kurtosis is computed from the accumulated power and power-squared, for both observing frequencies, using Equation \ref{skestimator}.  From the kurtosis, the kurtstrum is computed on hour-long timescales, using the algorithm described in Equation \ref{kurtstrum}.  Because we expect relevant signals in the kurtstrum to be strong enough that they are readily apparent, visual inspection, rather than automated thresholding, of the kurtstrum on hour-long timescales is used to identify areas of potential interest.

The wideband non-thermal microwave emissions that characterize electrostatic discharge \citep{Renno:2003vk} are expected to appear in the spectral kurtosis as significant deviations in the spectral kurtosis from the average Gaussian value over a wide range of frequency channels \citep{Ruf:2009uy}.  Low-frequency modulation of the non-thermal emission (caused by processes described in Section \ref{discussion}) would appear as spectral peaks in the kurtstrum at the fundamental modes of Mars' Schumann resonance.  In particular, we expect to see similar variations at 3.2 and 8.0 GHz from broadband continuum emission.

\subsection{Mars Observations from 23 March 2010} \label{analysis23}
\begin{figure}[h!]
\begin{center}
\subfigure [Band 1: 3226.2 - 3252.4 MHz] {\label{completekurt1} \includegraphics[width=.485\textwidth]
  {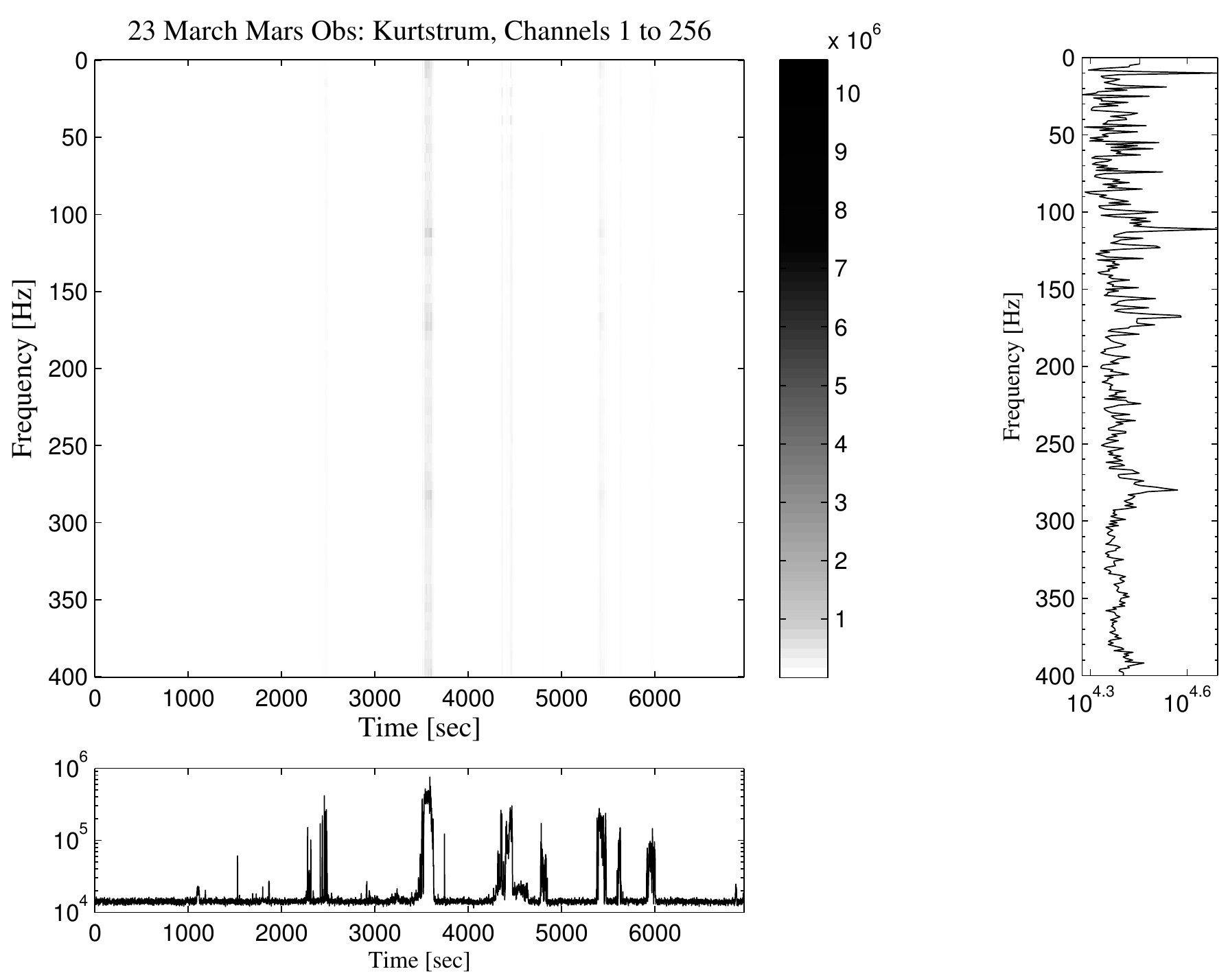}}
\subfigure [Band 2: 3200 - 3226.2 MHz] {\label{completekurt2} \includegraphics[width=.485\textwidth]
  {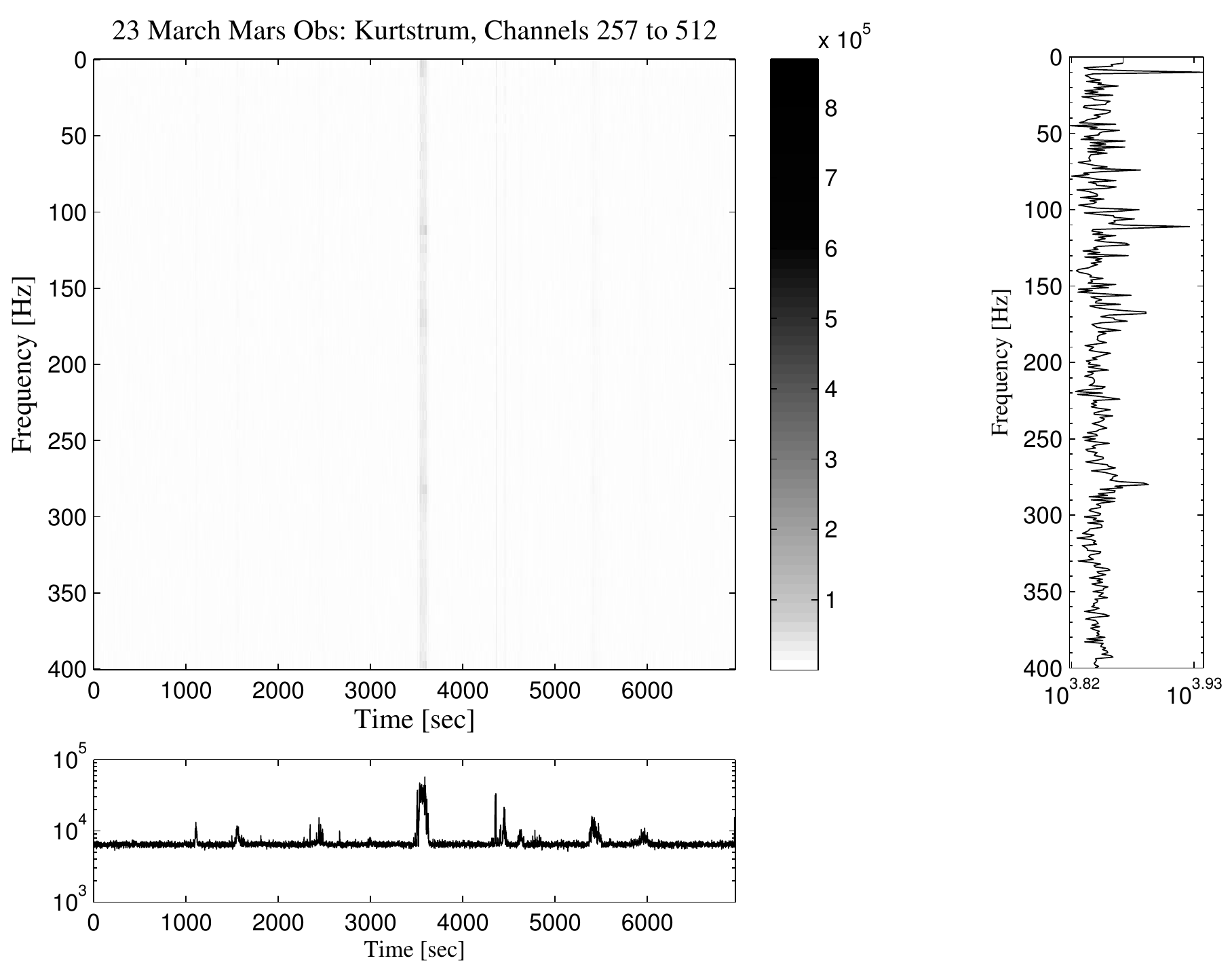}}
\end{center}
\begin{center}
\subfigure [Band 3: 3173.8 - 3200 MHz] {\label{completekurt3} \includegraphics[width=.485\textwidth]
  {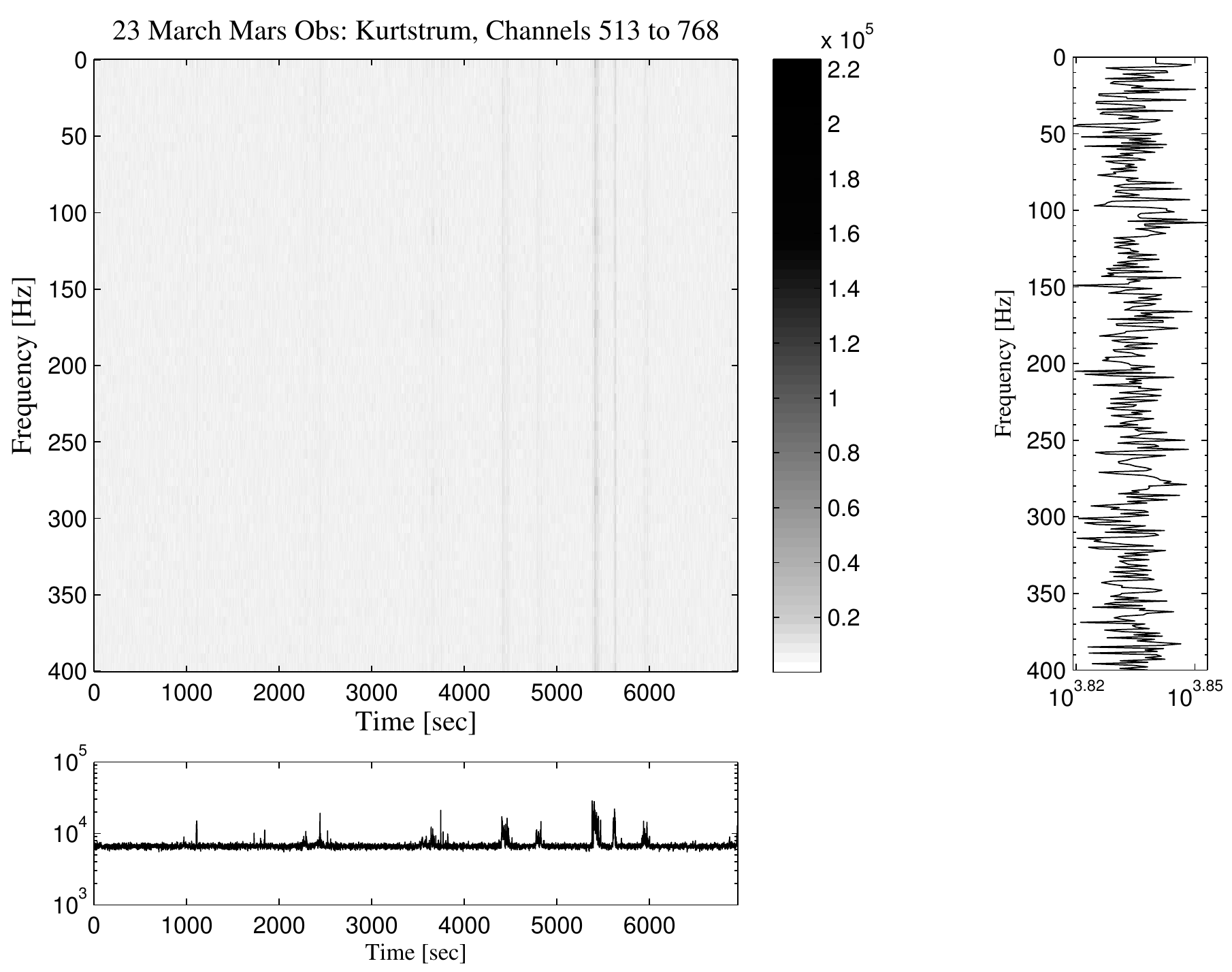}}
\subfigure [Band 4: 3147.6 - 3173.8 MHz] {\label{completekurt4} \includegraphics[width=.485\textwidth]
  {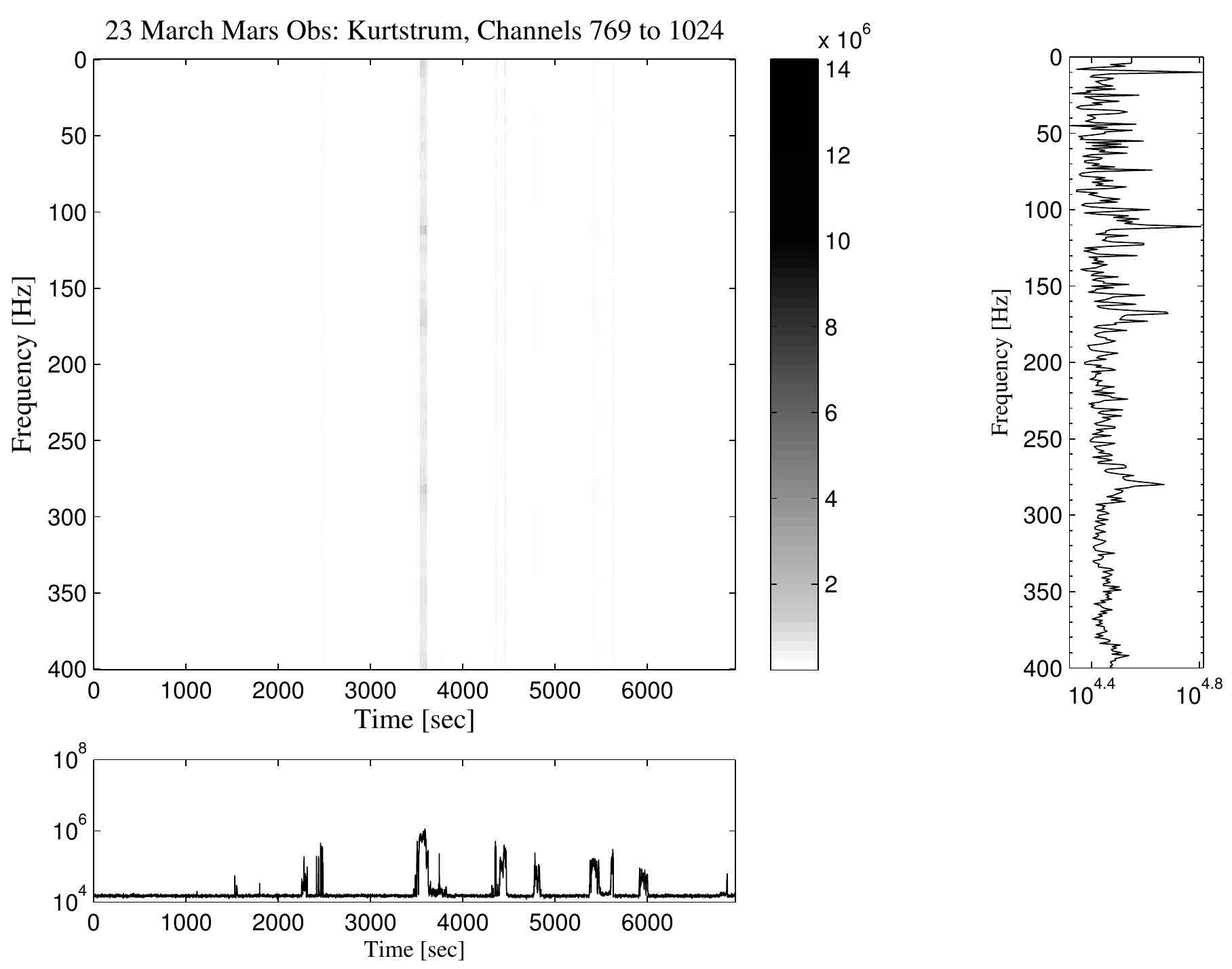}}
\end{center}
\caption{Kurtstrum (at 3.2 GHz, y-polarization) for Mars observations on 23 March 2010, spanning approximately two hours, from 01:11$-$03:07 UTC. The kurtstrum is computed by collapsing the 1024 frequency channels of the kurtosis into four bands, and computing the spectrum for each one second of data, giving a 1 Hz frequency resolution.  Figures (a-d) show the kurtstrum for all four bands, as a function of frequency on the y-axis and elapsed observation time on the x-axis.  The plot to the right of each gray-scale plot is the time-averaged spectrum; the plot directly below each gray-scale plot is the spectrum averaged over frequency.  The time-averaged spectra show the signal peaking at approximately 10 Hz and higher order harmonics.  Multiple interference events are visible in all four bands; the focus in Figures \ref{zoomkurt} and \ref{zoomkurtbf2} is on the event at 3500 seconds, which is prominent at 3.2 GHz in the first, second, and fourth bands, but is present at 8.0 GHz in the first band only. \label{completekurt}}
\end{figure}

\begin{figure}[h!]
\begin{center}
\subfigure [Band 1: 3226.2 - 3252.4 MHz] {\label{zoomkurt1} \includegraphics[width=.485\textwidth]
  {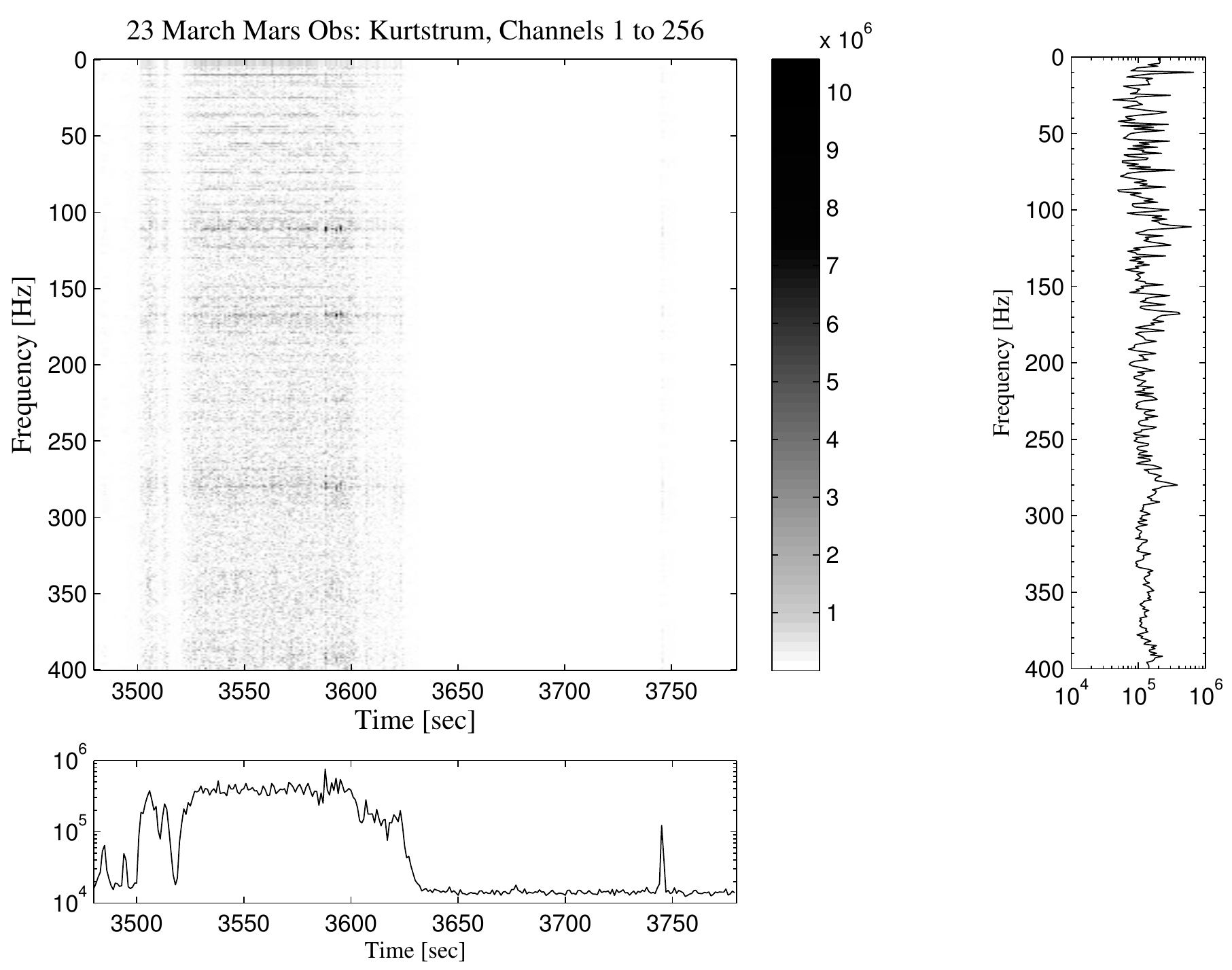}}
\subfigure [Band 2: 3200 - 3226.2 MHz] {\label{zoomkurt2} \includegraphics[width=.485\textwidth]
  {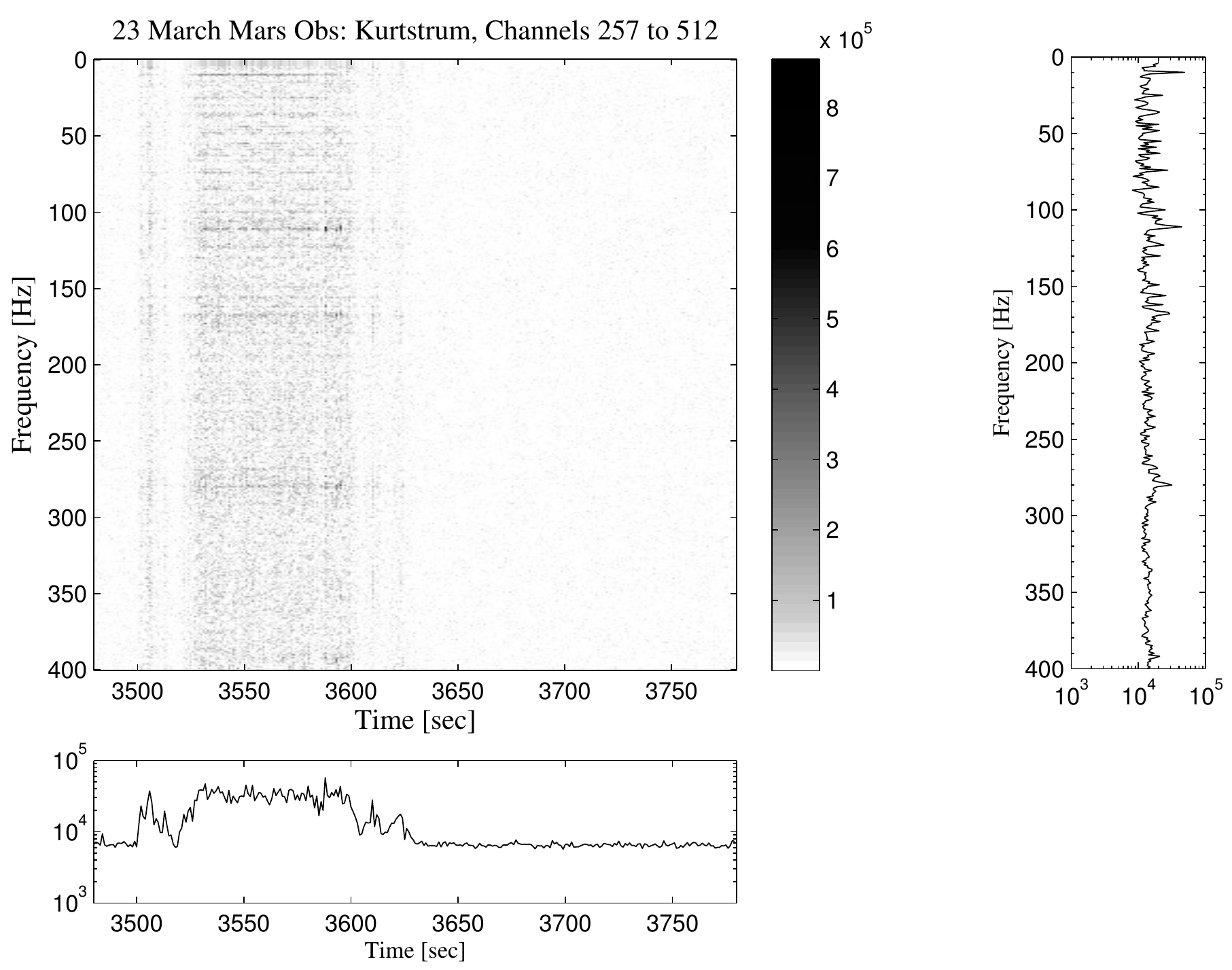}}
\subfigure [Band 3: 3173.8 - 3200 MHz] {\label{zoomkurt3} \includegraphics[width=.485\textwidth]
  {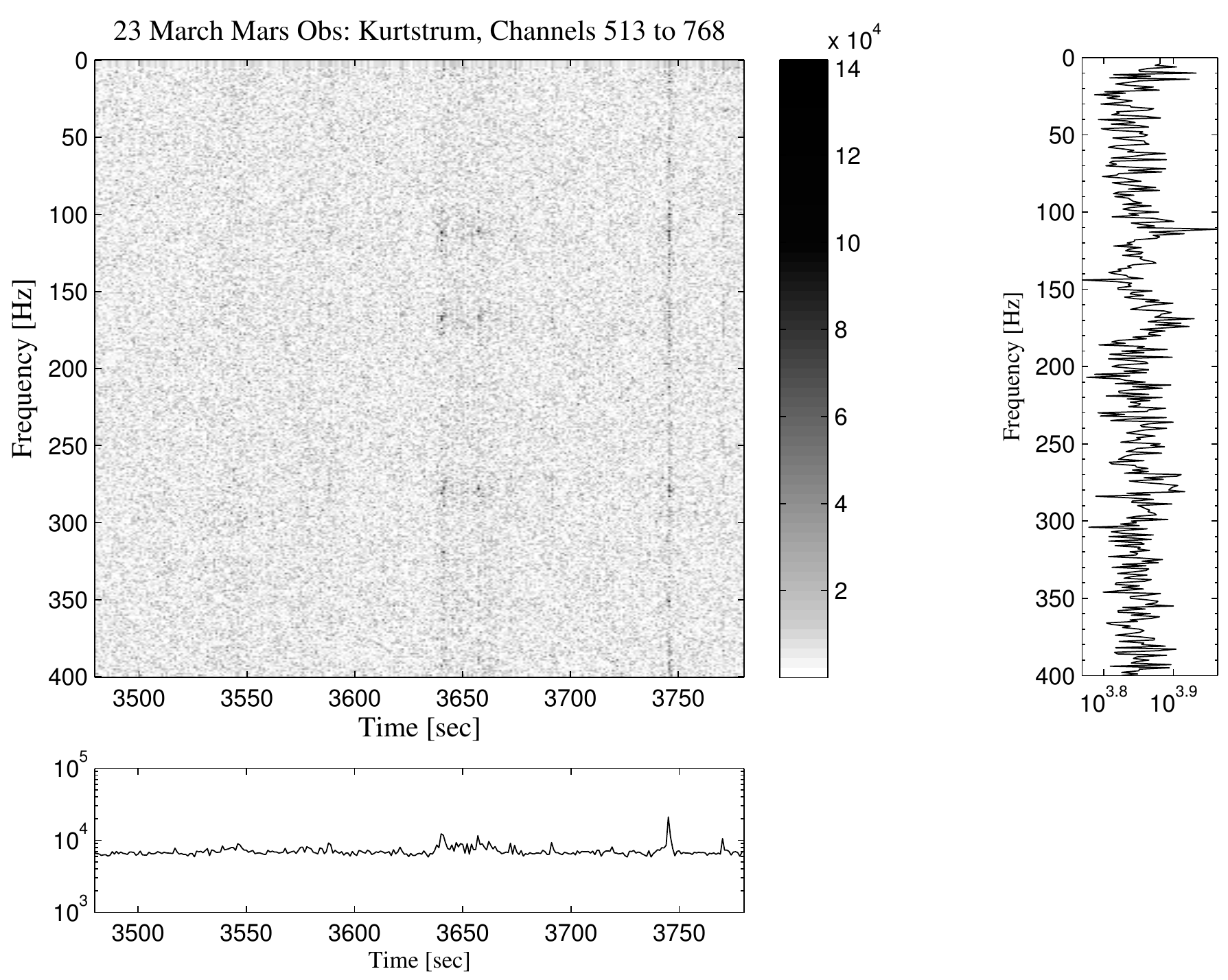}}
\subfigure [Band 4: 3147.6 - 3173.8 MHz] {\label{zoomkurt4} \includegraphics[width=.485\textwidth]
  {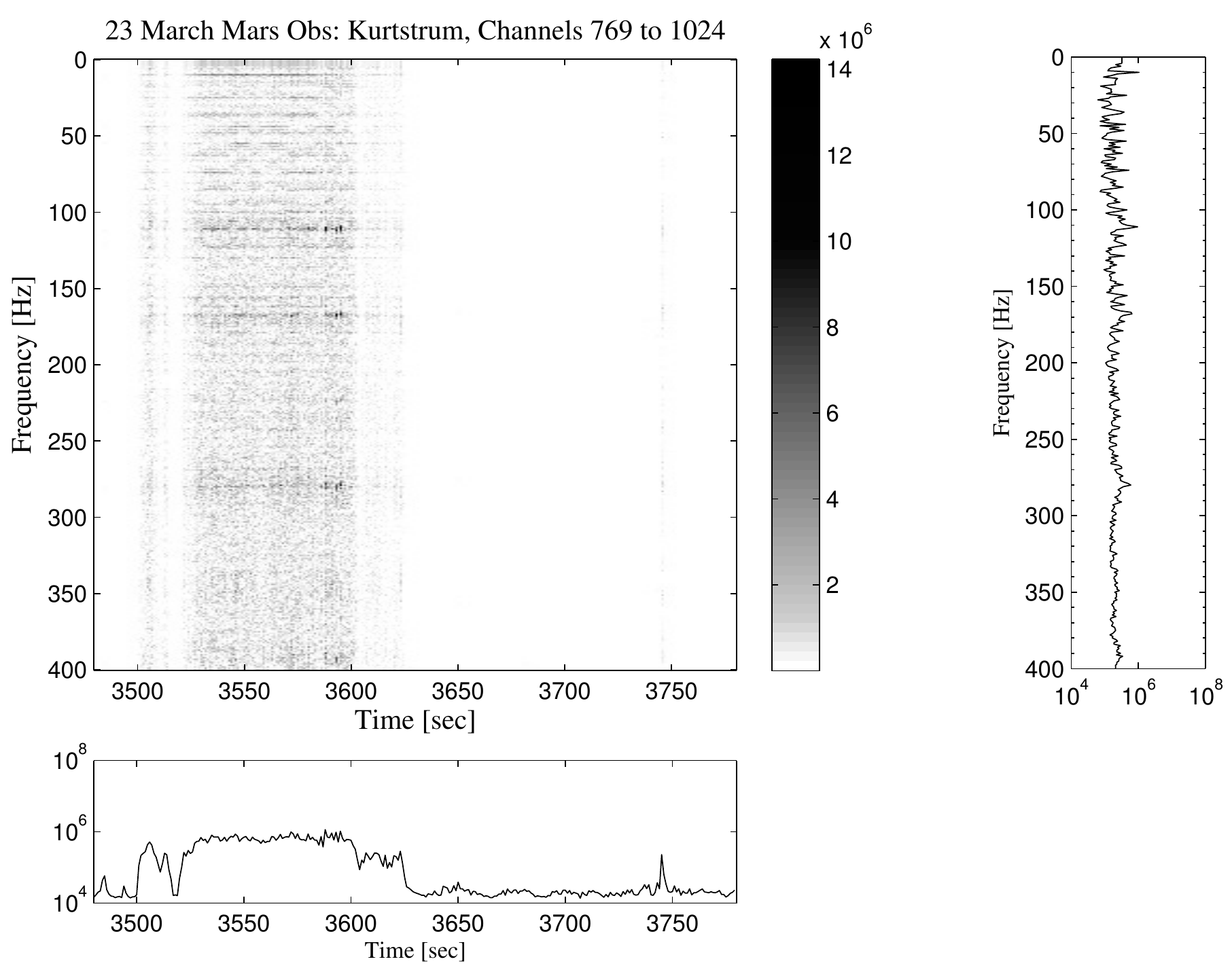}}
\end{center}
\caption{Kurtstrum (at 3.2 GHz, y-polarization) from 23 March 2010 over a period of five minutes, highlighting the event visible at 3500 seconds in Figure \ref{completekurt} at approximately 02:10 UTC. \label{zoomkurt}}
\end{figure}

We examine a characteristic set of data from 23 March 2010, beginning with our observations at 3.2 GHz and following with those at 8.0 GHz.  An example of the kurtstrum over a two-hour long period, from observations of Mars at 3.2 GHz on 23 March 2010 between 01:11--03:07 UTC, is given in Figure \ref{completekurt} for the four bands over which the kurtstrum was calculated.  Multiple events are visible in the kurtstrum, often occurring simultaneously in more than one or all bands at 3.2 GHz.  Figure \ref{zoomkurt} shows one of these events for a five-minute period of observation for all four bands at 3.2 GHz.  At this time-scale, the duration and frequency structure of the event in the kurtstrum is clearer, specifically the strong peak at 10 Hz and subsequent peaks at approximately 10 Hz harmonics in the first, second, and fourth bands.

\begin{figure}[h!]
  \begin{center}
    \subfigure [] {\label{ypower} \includegraphics[width=.9\textwidth]{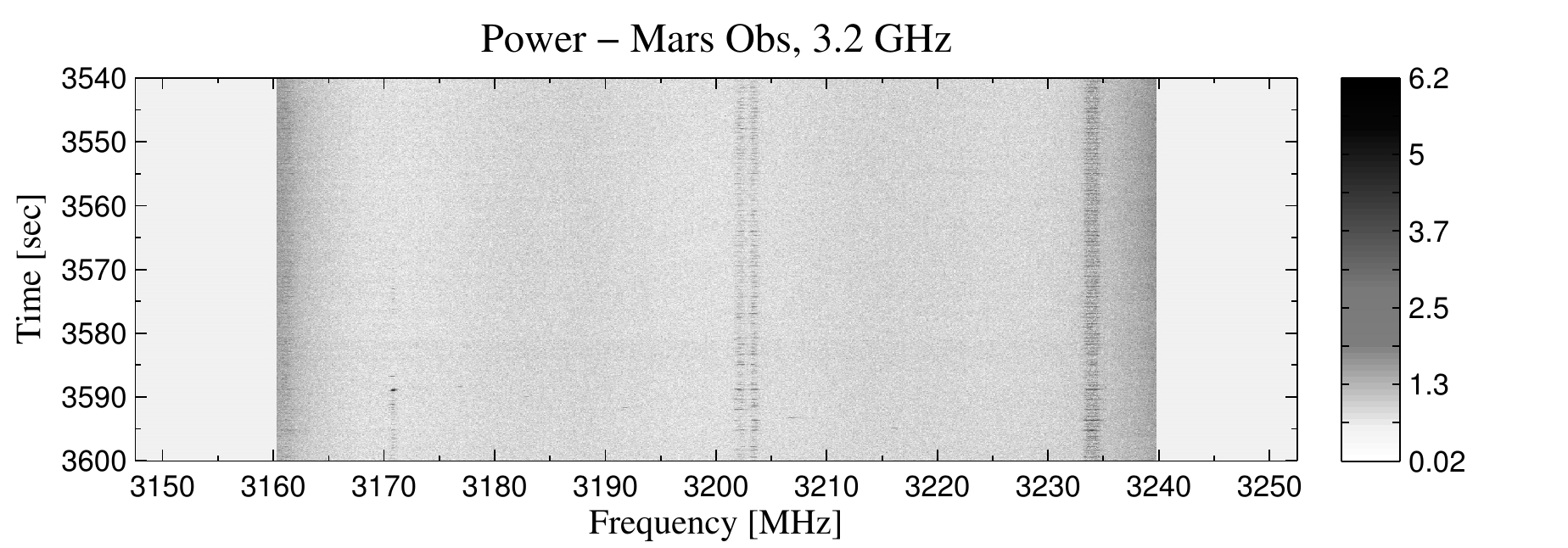}}
  \end{center}
  \begin{center}
    \subfigure [] {\label{ykurtosis} \includegraphics[width=.9\textwidth]{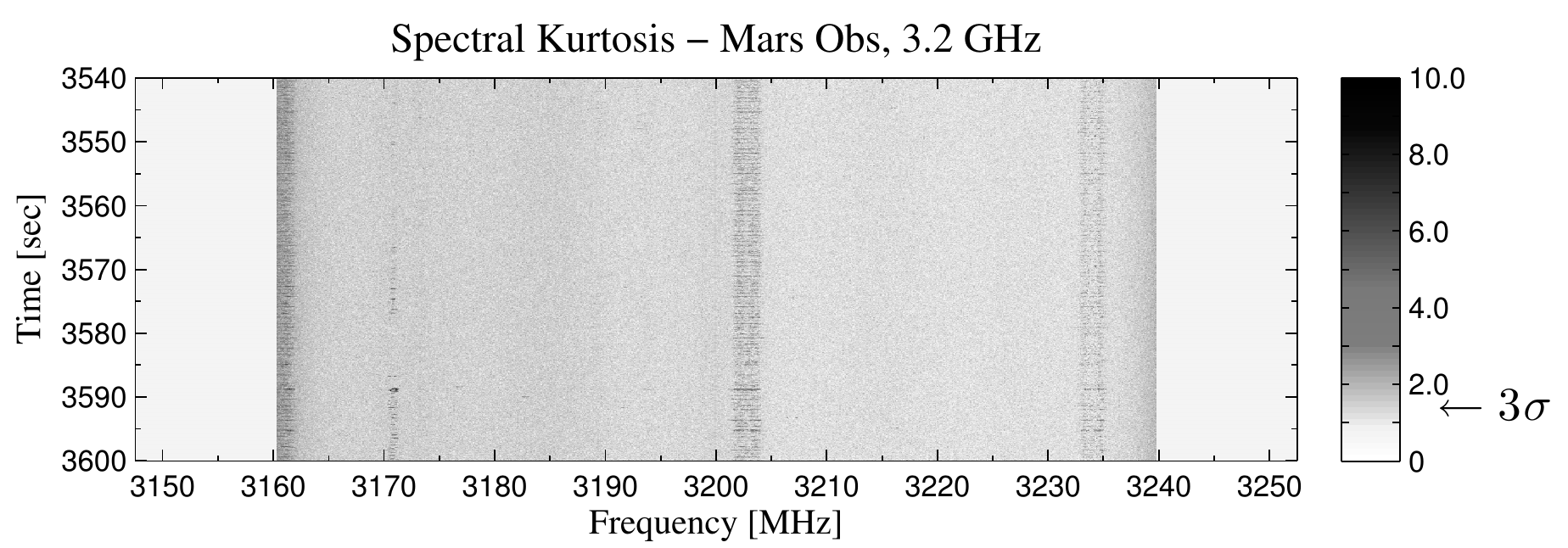}}
  \end{center}
  \caption{Power and spectral kurtosis as a function of sky frequency and time at 3.2 GHz, y-polarization, over the 60-second period during which narrowband interference caused the strong, $\sim$10 Hz modulated signal in the kurtstrum, during the 23 March 2010 Mars observations.  Interference is visible in all three over multiple frequency channels.  The $3\sigma$ kurtosis detection threshold is indicated in the colorbar on the right.  The band edges in the plots have been removed due to poor instrument response in those regions. \label{total}}
\end{figure}

\begin{figure}[h!]
  \begin{center}
    \subfigure [Spectral kurtosis: 3200 - 3206.5 MHz] {\label{kurtrfi} \includegraphics[height=.4\textheight]
{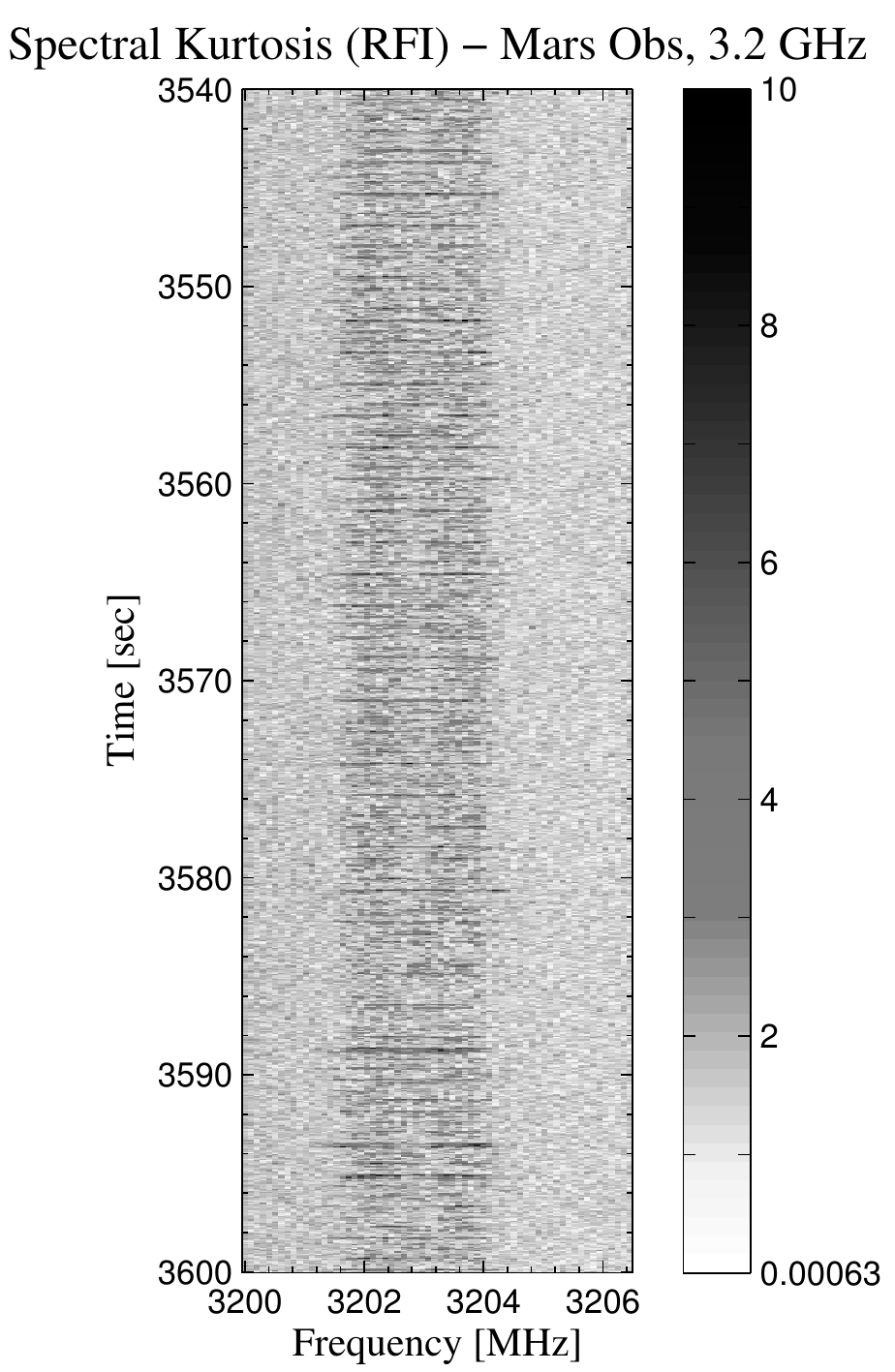}}
    \subfigure [Kurtstrum: 3200 - 3206.5 MHz] {\label{kurtstrumrfi} \includegraphics[width=.6\textwidth]
{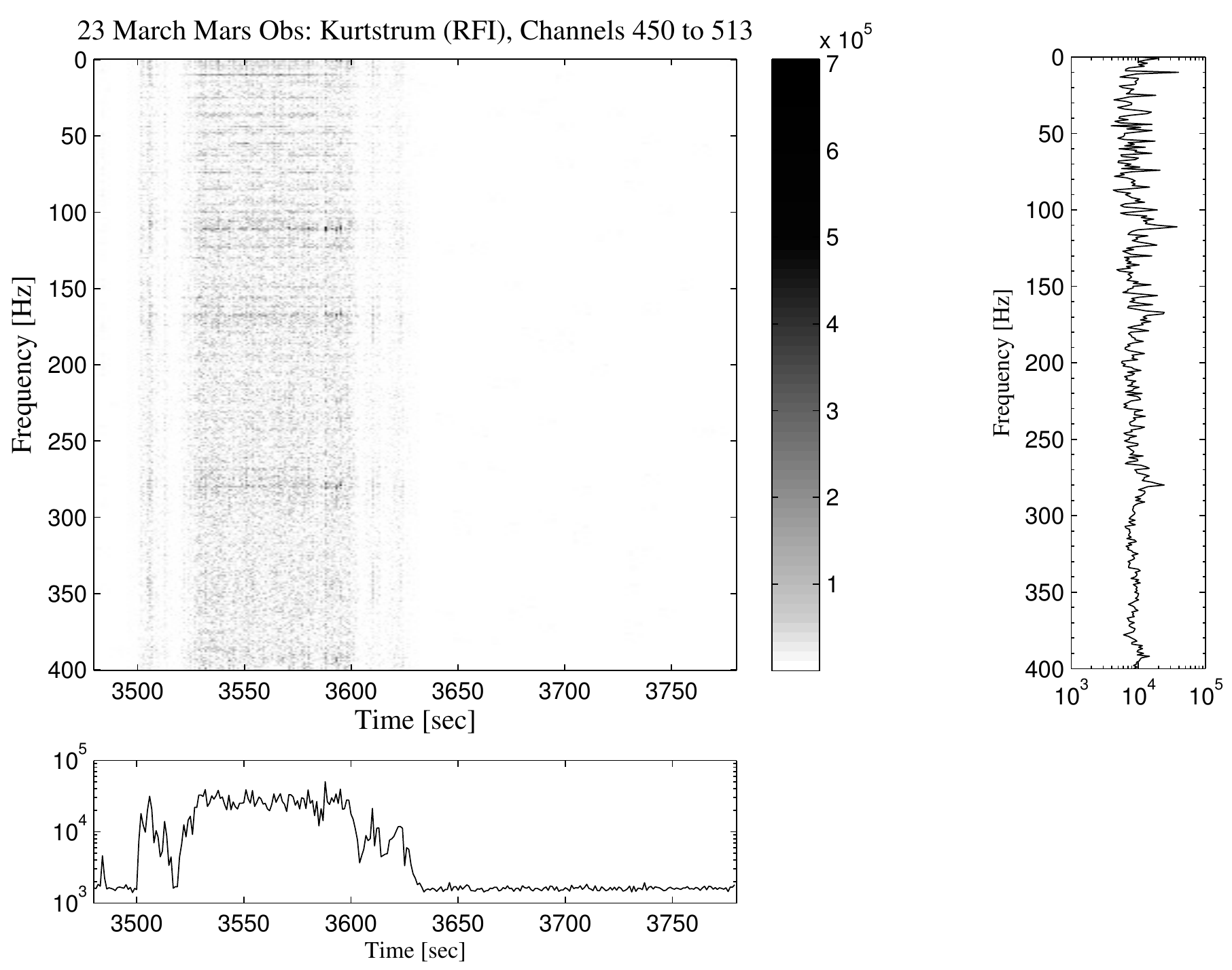}}
  \end{center}
  \begin{center}

    \subfigure [Spectral kurtosis: 3206.5 - 3213 MHz] {\label{kurtnorfi} \includegraphics[height=.4\textheight]
{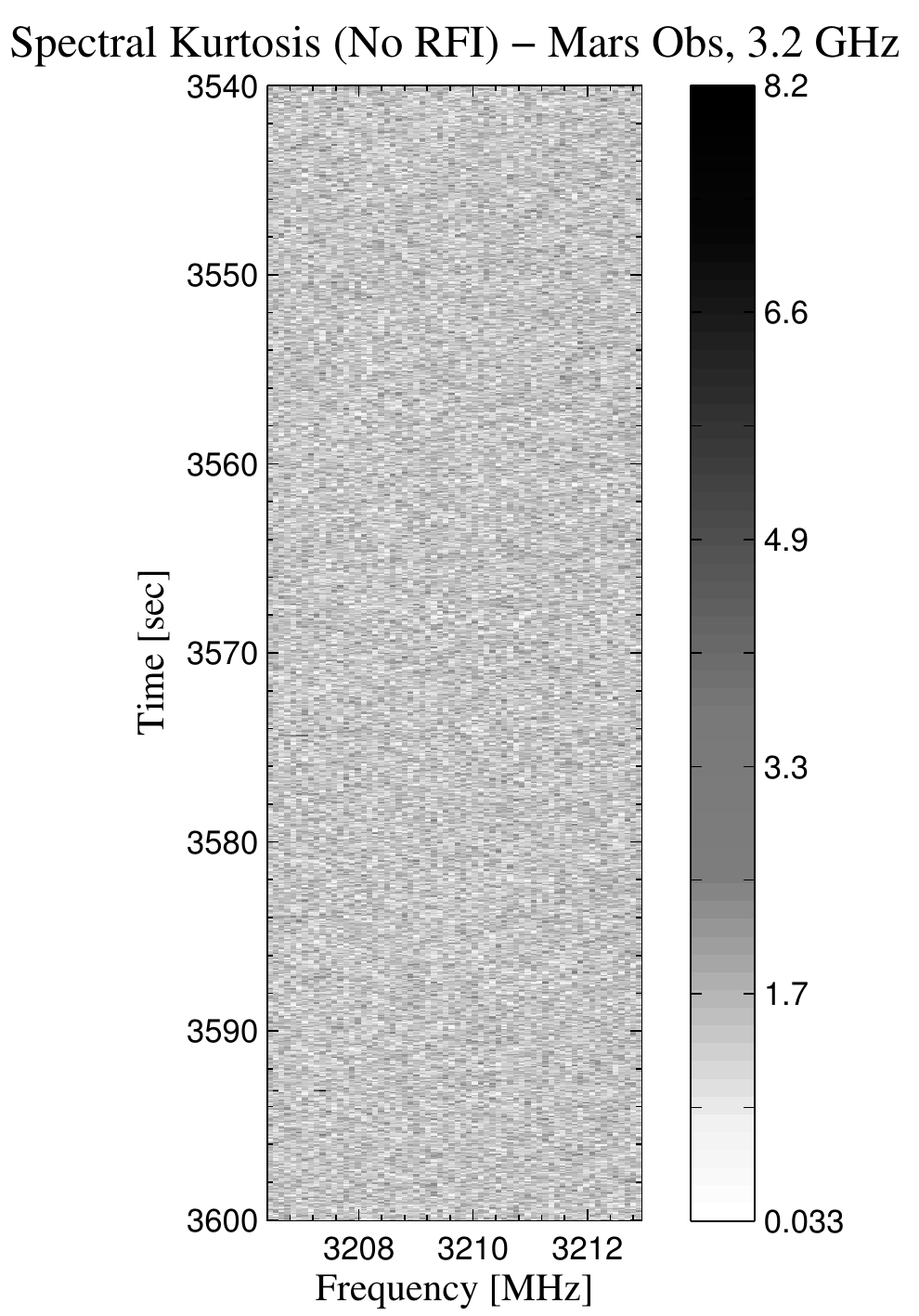}}
    \subfigure [Kurtstrum: 3206.5 - 3213 MHz] {\label{kurtstrumnorfi} \includegraphics[width=.6\textwidth]
{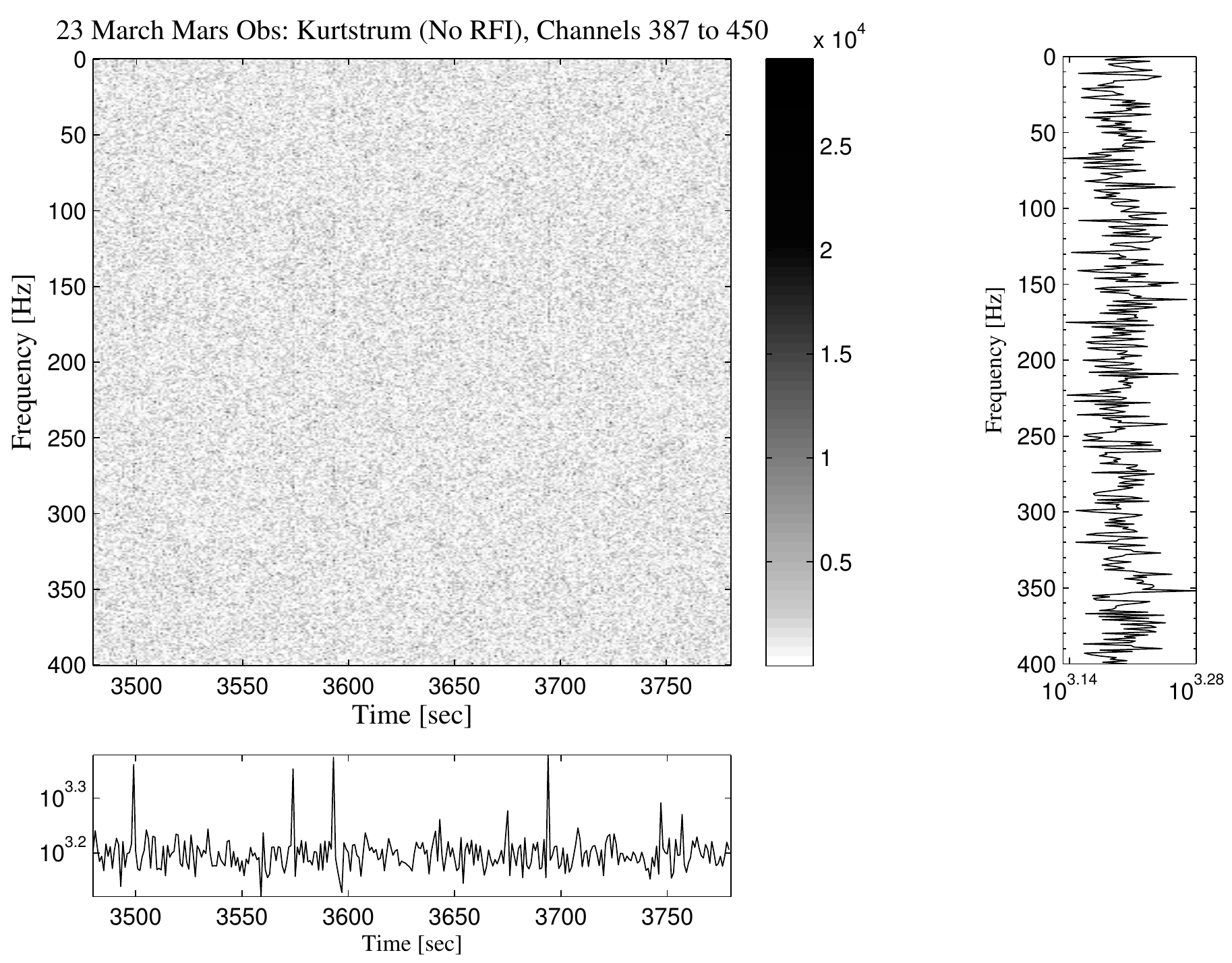}}
  \end{center}
  \caption{The spectral kurtosis at 3.2 GHz (a) from Figure \ref{ykurtosis}, confined to a region where narrowband RFI is present.  The computed kurtstrum (b) for this 6.5 MHz-wide band peaks during the two-minute period in which the RFI is present, and contains the characteristic 10 Hz and higher harmonics spectral features.  The spectral kurtosis (c), also from Figure \ref{ykurtosis}, confined to a 6.5 MHz-wide band adjacent to (a) in which there is no RFI.  Its corresponding kurtstrum (d) contains no 10 Hz spectral features. \label{narrow}}
\end{figure}

Power across a wide frequency range with 10 Hz variations in the kurtstrum are consistent with the expectations for electrostatic discharges as described by \citet{Ruf:2009uy}.  A closer look, however, reveals that the variability we see is driven by narrowband features that are spread across a broad spectrum.  A plot of the spectral kurtosis and corresponding accumulated power in Figure \ref{total} from which the kurtstrum at 3.2 GHz in Figure \ref{zoomkurt} is calculated shows a constant elevation of kurtosis levels above the 3$\sigma$ detection threshold over narrow ranges of frequency channels for the duration of the event in the kurtstrum.  Dividing the kurtosis into four bands (from right to left), it is evident that this narrowband RFI is present in precisely the bands that show the strongest 10 Hz-modulated signals in the kurtstrum (namely, the first, second, and fourth bands ranging from 3226.2 - 3252.4 MHz, 3200 - 3226.2 MHz, and 3147.6 - 3173.8 MHz, respectively).  The effect of narrowband RFI on the kurtstrum is more clearly demonstrated in Figure \ref{narrow}, which shows the kurtstrum computed for two adjacent 6.5 MHz-wide bands, the first with strong RFI and the second without.  The kurtstrum computed from the spectral kurtosis band containing RFI is nearly identical to Figures \ref{zoomkurt1}, \ref{zoomkurt2}, \ref{zoomkurt4}, and exhibits the characteristic 10 Hz and higher harmonics frequency structure.

\begin{figure}[h!]
\begin{center}
\subfigure [Band 1: 8026.2 - 8052.4 MHz] {\label{zoomkurt1bf2} \includegraphics[width=.485\textwidth]
  {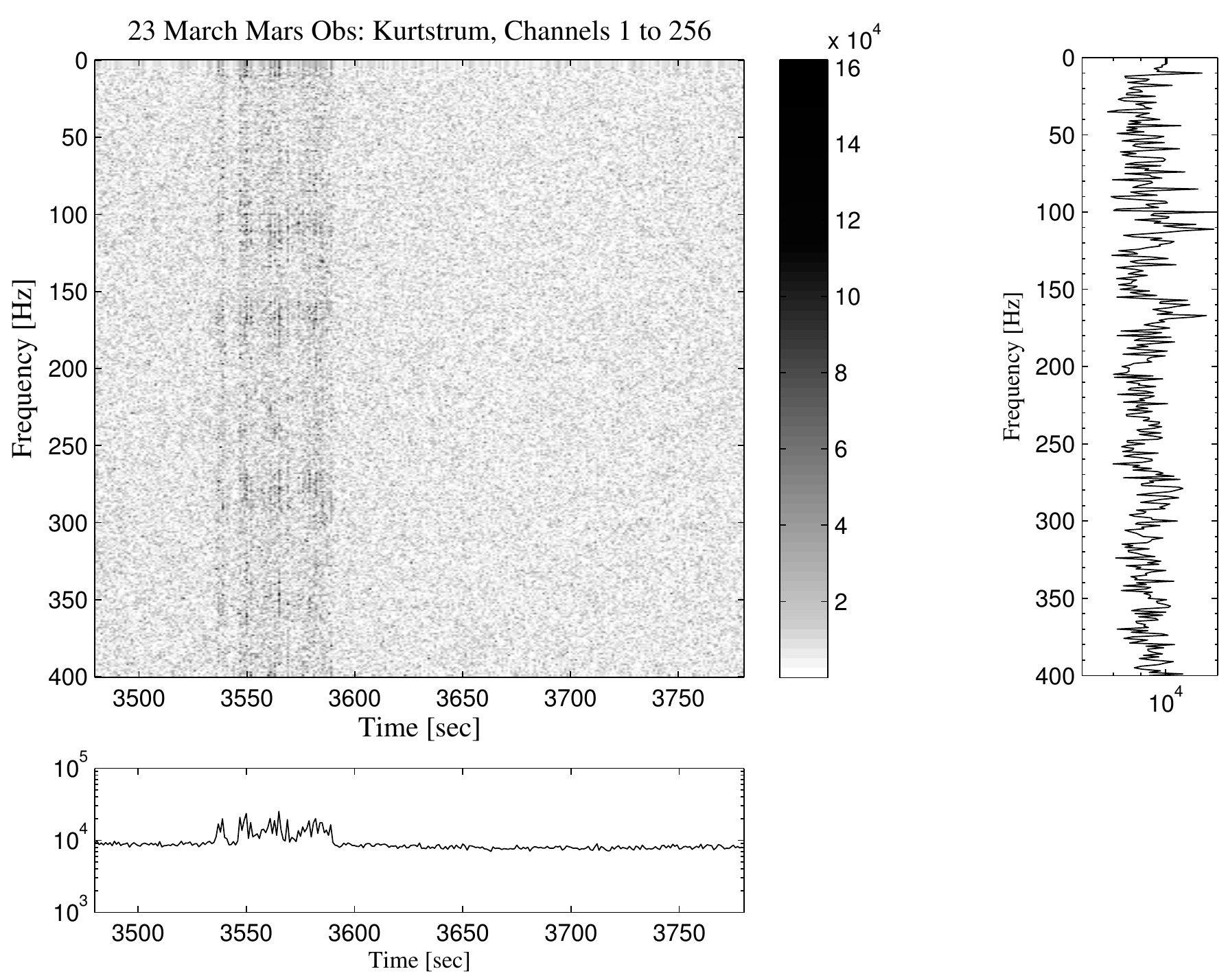}}
\subfigure [Band 2: 8000 - 8026.2 MHz] {\label{zoomkurt2bf2} \includegraphics[width=.485\textwidth]
  {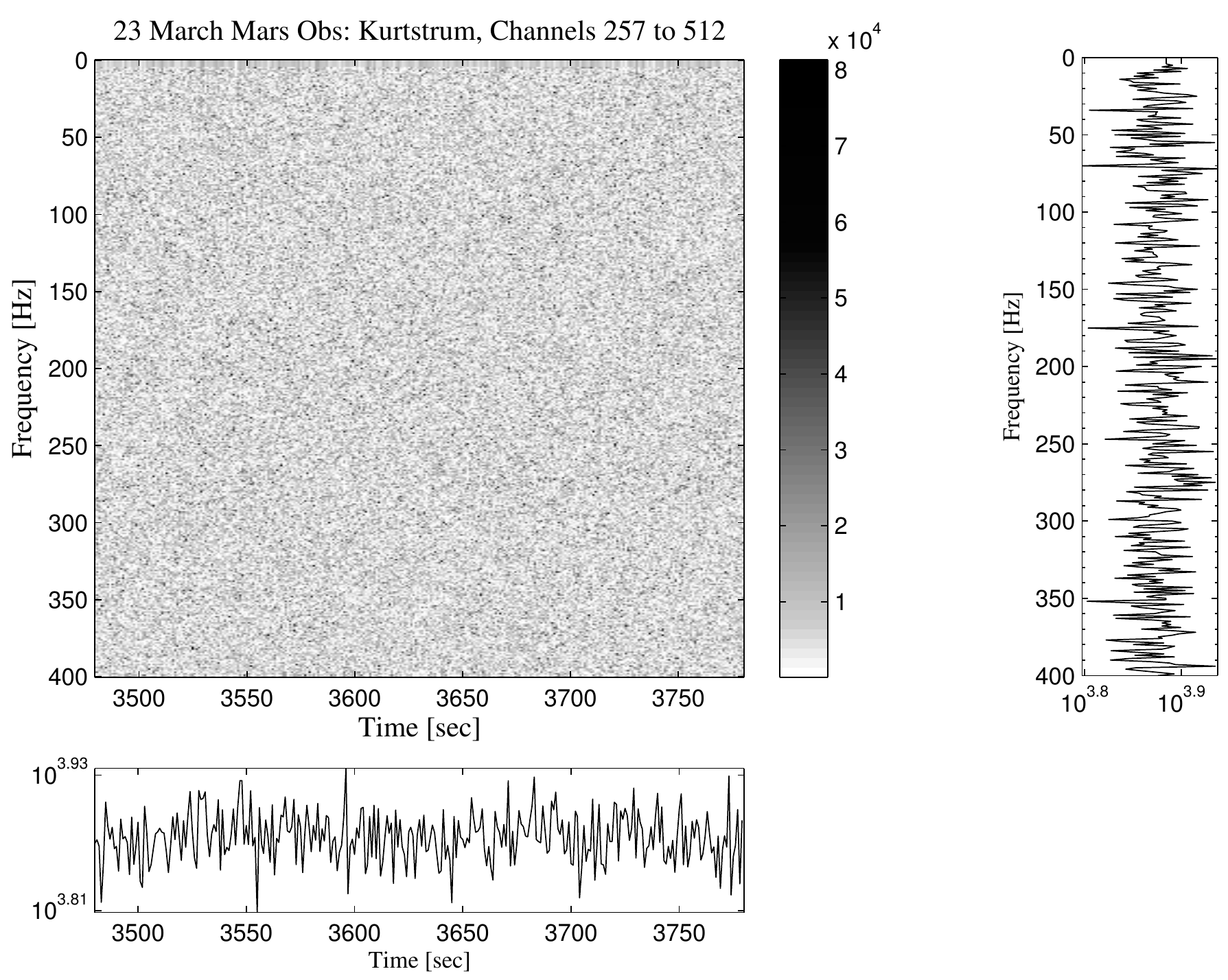}}
\subfigure [Band 3: 7973.8 - 8000 MHz] {\label{zoomkurt3bf2} \includegraphics[width=.485\textwidth]
  {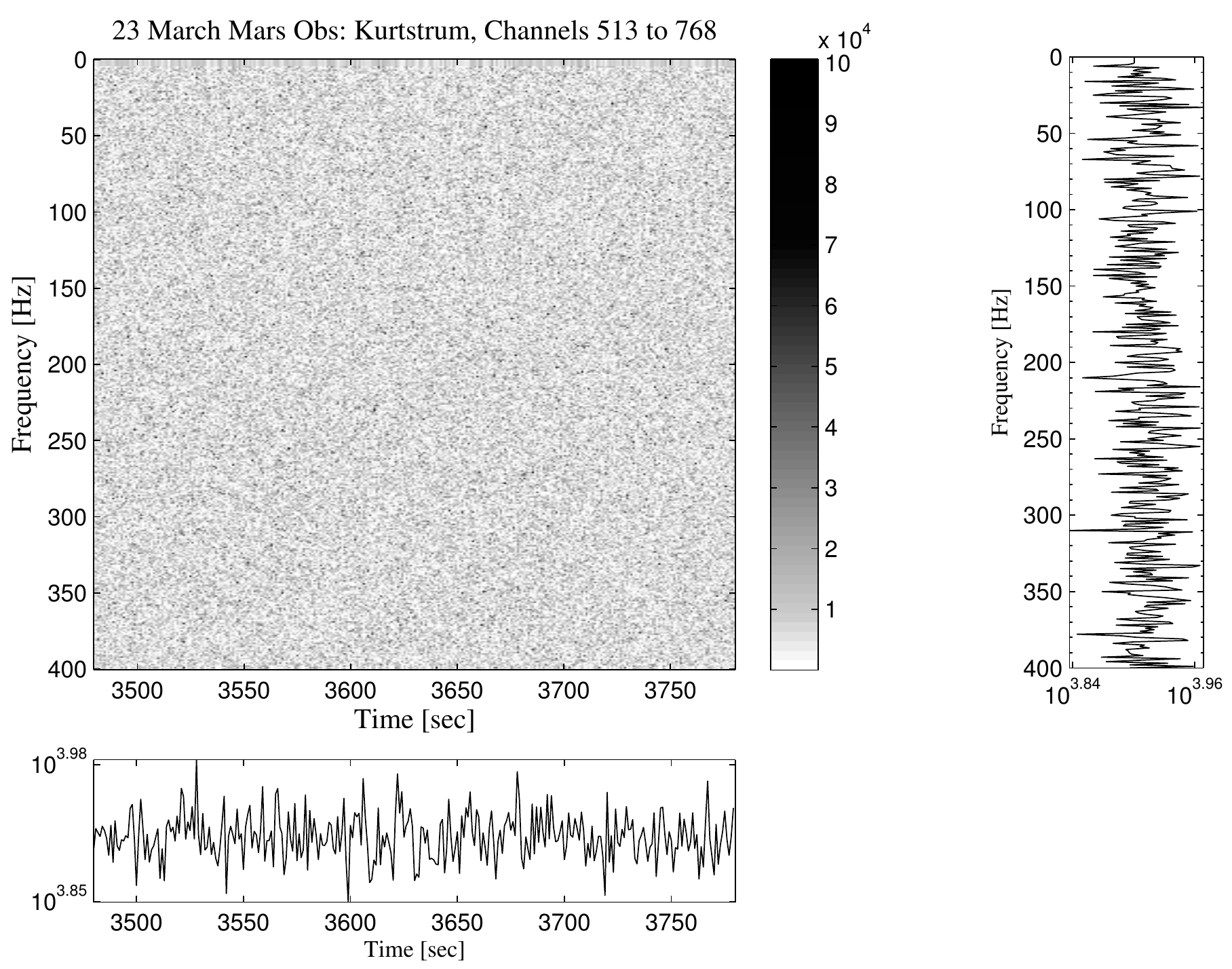}}
\subfigure [Band 4: 7947.6 - 7973.8 MHz] {\label{zoomkurt4bf2} \includegraphics[width=.485\textwidth]
  {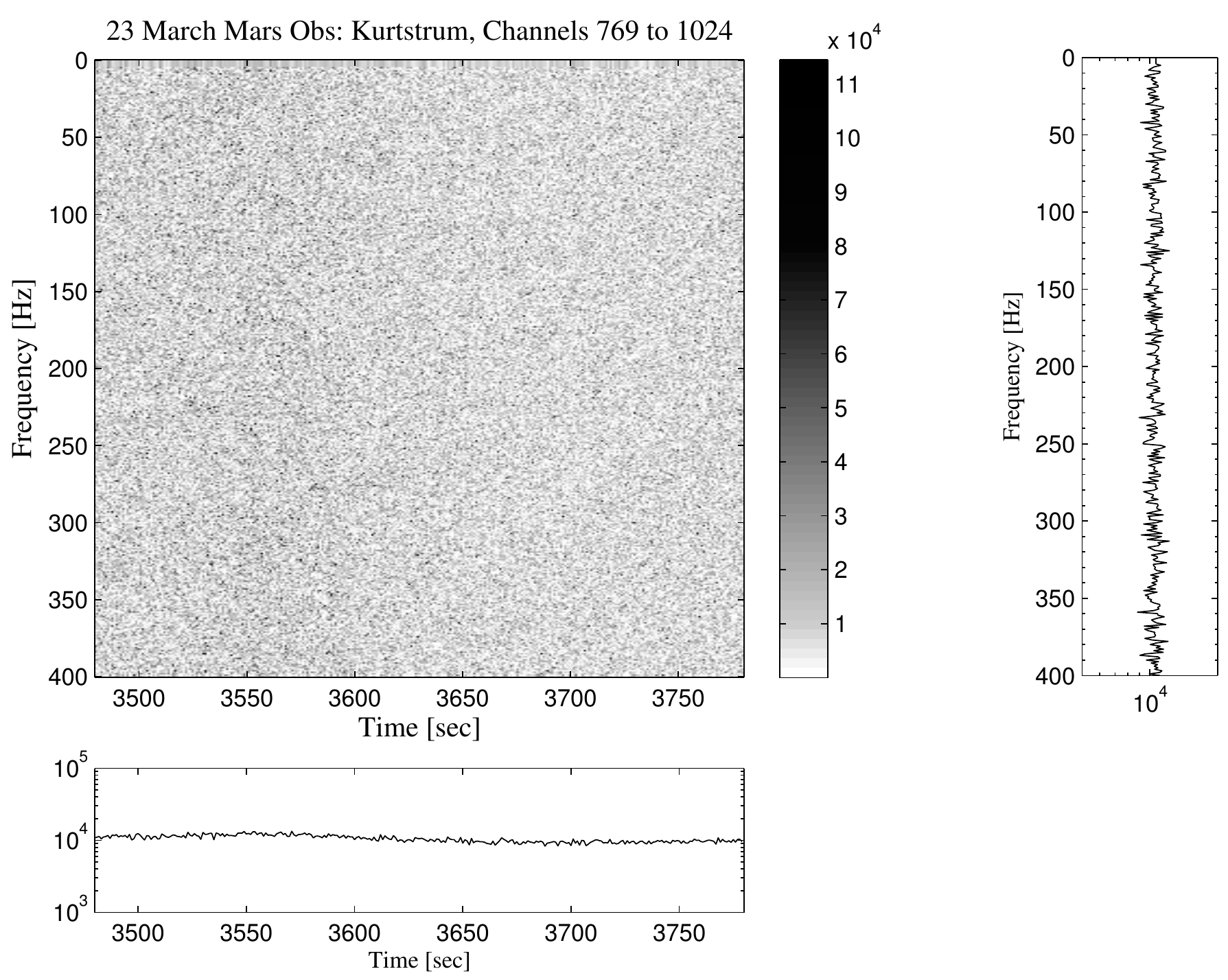}}
\end{center}
\caption{Kurtstrum, corresponding to Figure \ref{zoomkurt}, at 8.0 GHz, y-polarization, over a period of five minutes.  The signal is visible only in the first band. \label{zoomkurtbf2}}
\end{figure}

\begin{figure}[h!]
\begin{center}
\subfigure [] {\label{tck1int} \includegraphics[width=.485\textwidth]{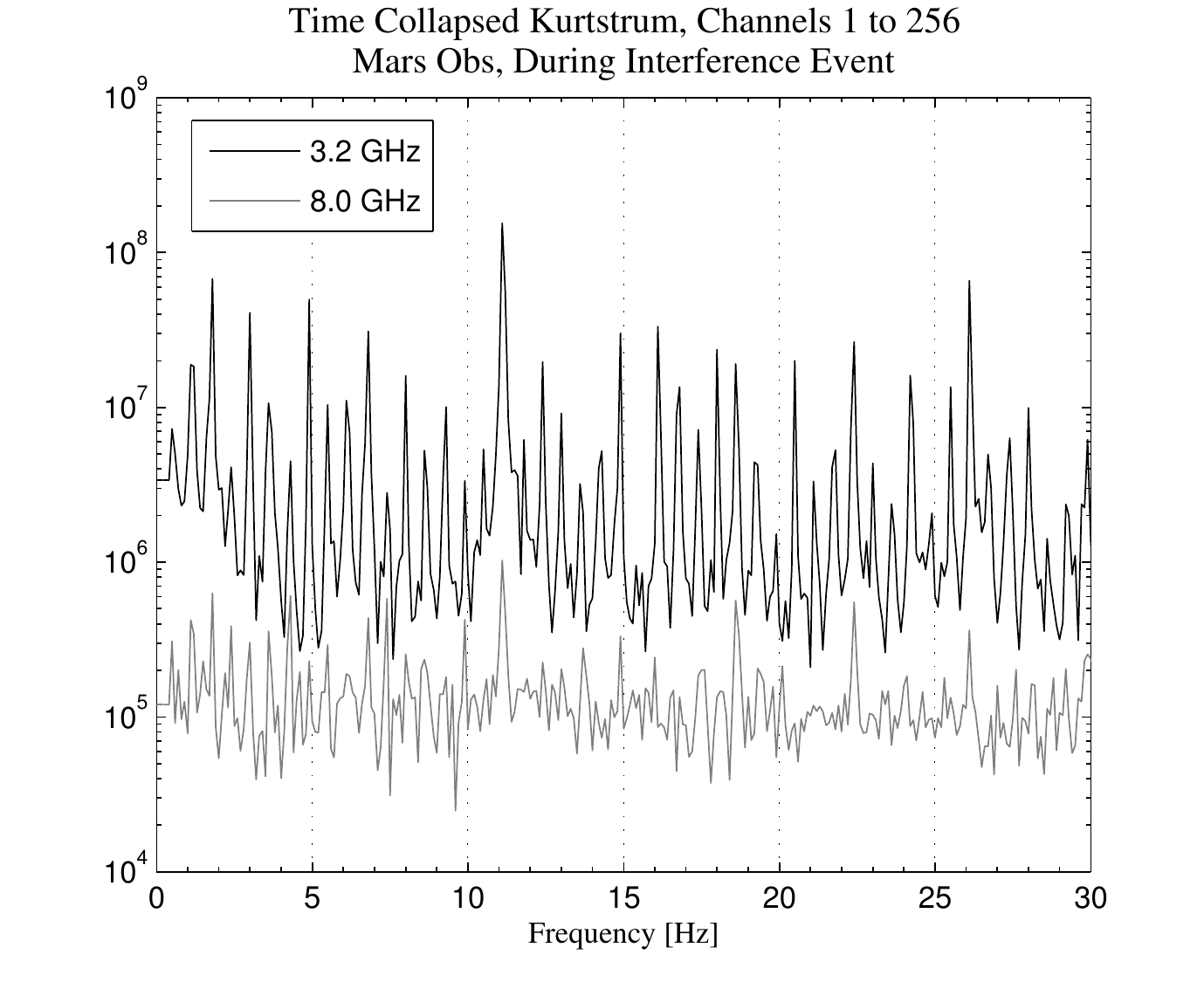}}
\subfigure [] {\label{tck1noint} \includegraphics[width=.485\textwidth]{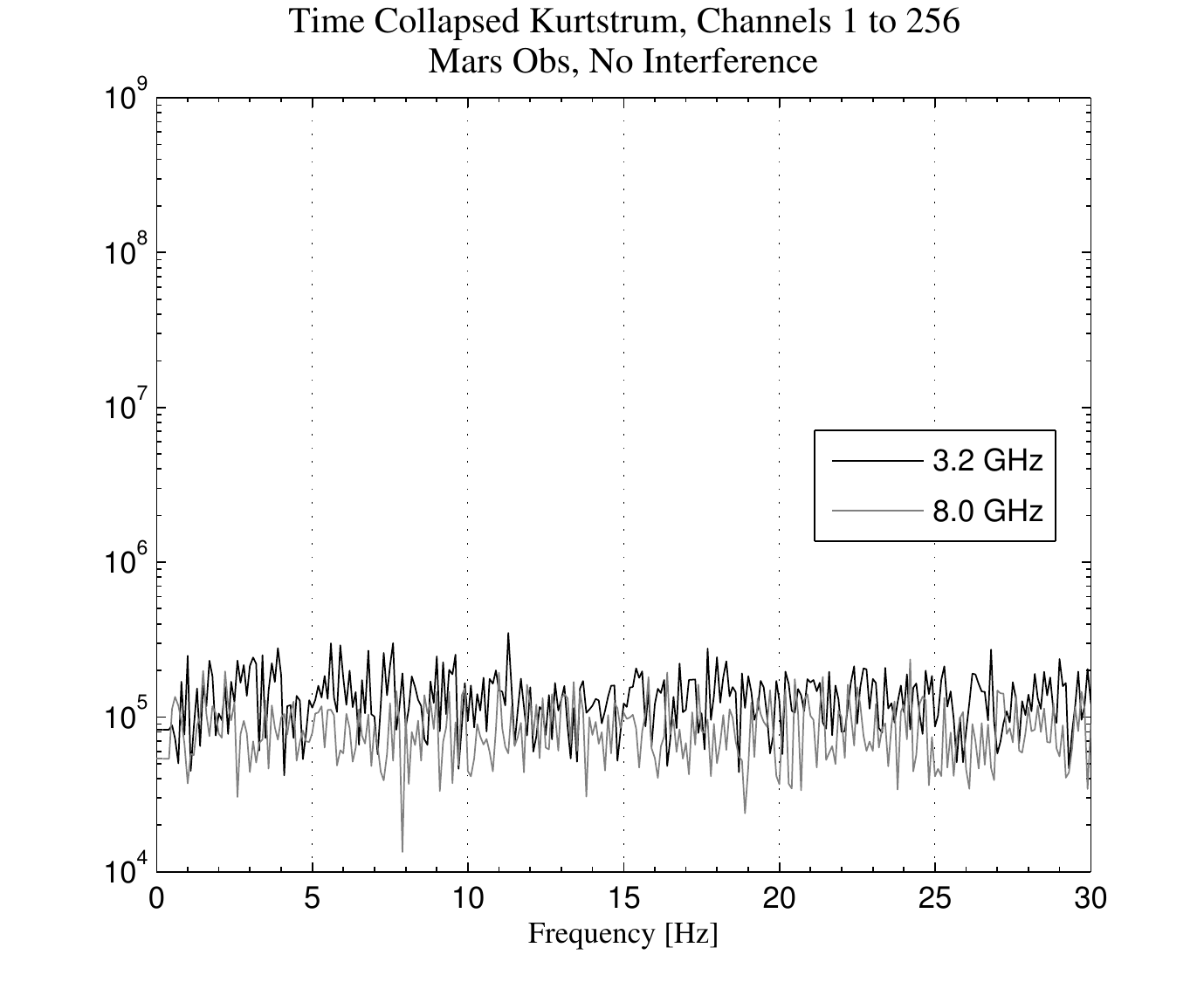}}
\caption{Time-averaged kurtstrum over a period of 60 seconds for the first 256 channels of both beamformers, at 3.2 and 8.0 GHz, on y-polarization, during an interference event (\ref{tck1int}) and during a period with no interference (\ref{tck1noint}).  The sharp spectral features at 3.2 and 8.0 GHz in \ref{tck1int} are correlated, and are caused by narrowband interference (visible in Figure \ref{total}) that is present in multiple frequency bands as higher harmonics or intermodulation products. \label{tck1}}
\end{center}
\end{figure}

At 8.0 GHz, 10 Hz variations are visible in the kurtstrum, similar to those seen at 3.2 GHz.  Figure \ref{zoomkurtbf2} shows the kurtstrum at 8.0 GHz over the same five-minute period shown in Figure \ref{zoomkurt}.  The strong signal with peaks at 10 Hz and higher harmonics that was present in three of the four 3.2-GHz bands is present, though significantly weaker, in the first 8.0-GHz band.  The signals in the kurtstrum at 3.2 and 8.0 GHz are highly correlated, suggesting a similar source of narrowband RFI for the signals at both frequencies.  We see even more clearly the high correlation between signals in the kurtstrum at 3.2 and 8.0 GHz (Figure \ref{tck1}) by narrowing the time range for analysis to 60 seconds.  Data selected from a period of no apparent interference show no structure in the kurtstrum.  The spectral kurtosis and corresponding accumulated power at 8.0 GHz, unlike at 3.2 GHz, contains no visible signs of narrowband RFI, due to a lack of sensitivity at this high frequency band.

The evidence then 
suggests that for this event on 23 March 2010 
the peaks in the kurtstrum at 3.2 and 8.0 GHz are caused by
narrowband  RFI that is spread over a broad frequency range.  
This is likely produced by strong RFI producing intermodulation products in the
analog signal path of the ATA.
The RFI is recognizable in the spectral kurtosis and power at 3.2 GHz as signals that are highly confined in frequency.  The bands at 3.2 GHz in which narrowband RFI are present are precisely those bands in which the 10 Hz-modulated signal appears in the kurtstrum at 3.2 GHz.  The RFI does not appear in the spectral kurtosis or power at 8.0 GHz, likely because the harmonics of the interferer at this higher frequency range are too weak to be detected.  However, its effects are visible in the kurtstrum at 8.0 GHz as the familiar 10 Hz-modulated signal, which correlates well with equivalent signal at 3.2 GHz.

\subsection{All Mars Observations} \label{analysisall}
\begin{figure}[h!]
\begin{center}
  \subfigure [] {\label{obsovertimebf1} \includegraphics[width = .85\textwidth]
    {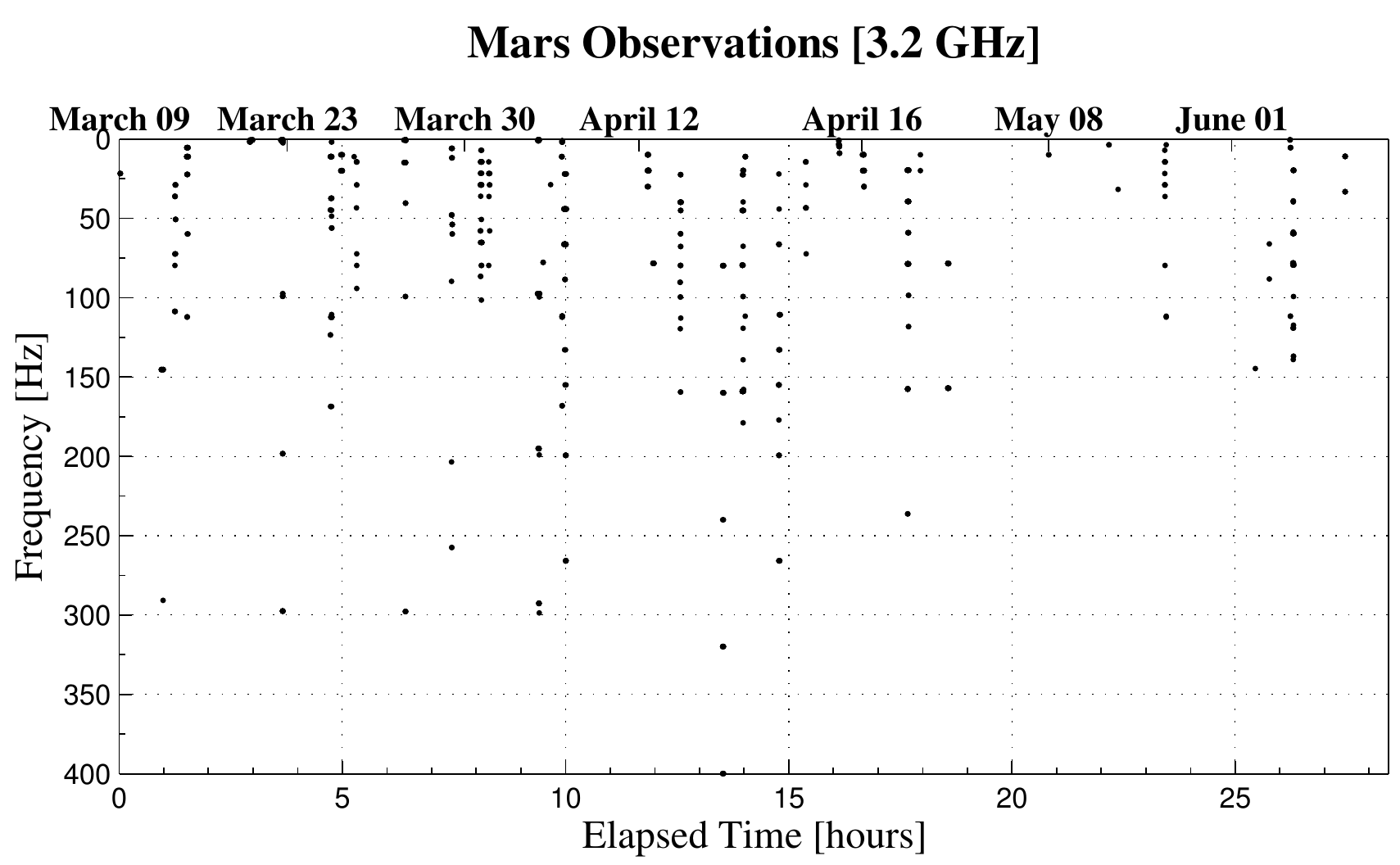}}
  \subfigure [] {\label{obsovertimebf2} \includegraphics[width = .85\textwidth]
    {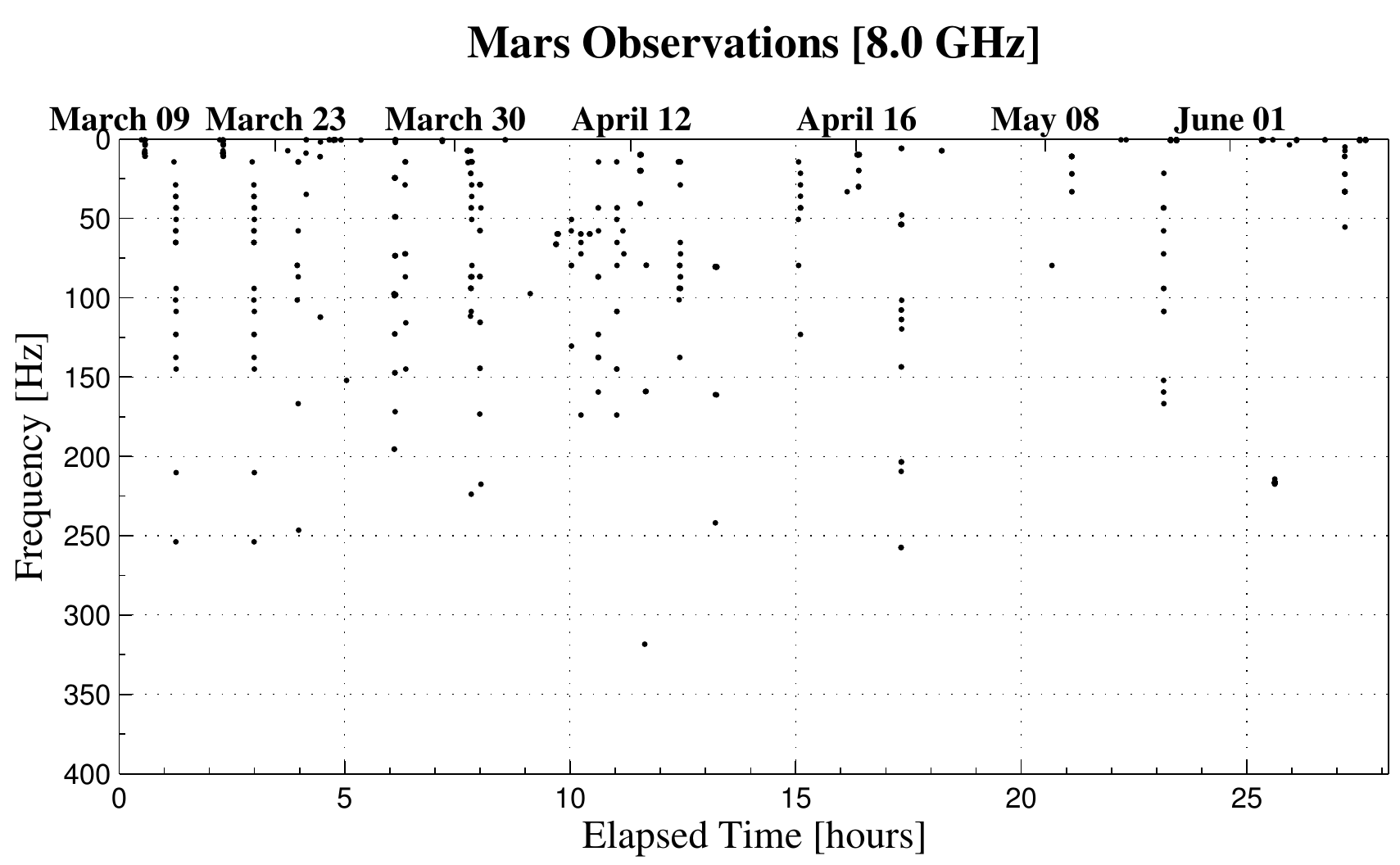}}
\end{center}
  \caption{Peak spectral frequencies of interference events in the kurtstrum versus their time of occurrence, at 3.2 GHz (top) and 8.0 GHz (bottom). \label{obsovertime}}
\end{figure}

The data from 23 March 2010 are qualitatively similar to events observed
from all 30 hours of Mars observations.  Significant events in the kurtstrum appear on average one to two times per hour of observation, and generally show spectral peaks near 10 Hz and higher order harmonics.  Figure \ref{obsovertime}, which plots the peak frequency of significant events in the kurtstrum for both 3.2 and 8.0 GHz against elapsed time for all 30 hours of Mars observations, demonstrates how often these events appeared in the kurtstrum, as well as the correlation between events at 3.2 and 8.0 GHz.  For these events with frequency structure resembling Martian Schumann resonance patterns, inspection of the corresponding spectral kurtosis and power (as a function of frequency and time) at 3.2 GHz shows the origin of the signals to be narrowband RFI, and not the wideband emission we would expect of lightning discharge.


\section{Discussion} \label{discussion}
\citet{Ruf:2009uy} argue that in the event of a lightning discharge in the Martian atmosphere, low-frequency electromagnetic waves are induced that propagate in the spherical cavity formed by the Martian surface and ionosphere; this sets off additional lightning discharges that occur in phase with the standing wave, similar to what occurs during terrestrial lightning storms.  The frequencies of these standing waves correspond to the fundamental modes of the Mars' Schumann resonance (SR).  These Schumann resonances are expected to appear as peaks in the power spectrum of the kurtosis \citep{Ruf:2009uy}, the lowest of which is predicted to fall in the range of 7-14 Hz \citep{Yang:2006wu}.  In the ideal case of a cavity whose ionosphere and surface boundaries are perfect conductors, the frequencies are given by:

\begin{equation} \label{schumanneq}
f_{n} = \frac{c}{2 \pi a} \sqrt{n (n+1)}
\end{equation}

\noindent where $c$ is the speed of light, $a$ is the radius of Mars, and $n$ is the mode number \citep{Schumann:1952}.  The measured terrestrial first SR mode is roughly 7.8 Hz \citep{Yang:2006wu}.  For Mars, the first SR mode given by Eq. \ref{schumanneq} is roughly 20 Hz, however realistic models incorporating conduction losses place the first resonance within the range of 7-14 Hz \citep{Yang:2006wu}.

Over the approximately 30 hours of observations on Mars, visual inspection of the data yielded no signals indicative of electrostatic discharge.  Repeated instances of strong signals in the kurtstrum that showed spectral peaks with a periodic 10 Hz structure were common in the Mars observations, appearing on the order of one to two times per hour of observations at 3.2 GHz (Figure \ref{completekurt}), and appearing less frequently at 8.0 GHz.  However, the spectral peaks were clearly harmonics and did not follow the resonant modes predicted by Schumann's equation (Equation~\ref{schumanneq}).  The strong correlation in frequency structure and time of occurrence in the kurtstrum between 3.2 and 8.0 GHz implies a common source for the signals, most likely appearing over a wide range of frequency bands as higher harmonics or intermodulation products from a single RFI source.  The frequency structure and timescales of the signals seen in the kurtstrum strongly resemble those detected by \citet{Ruf:2009uy} during the 8 June 2006 large-scale dust storm event -- spectral peaks occurring as higher harmonics of a fundamental near 10 Hz that are either close to or within the frequency ranges predicted for the first few fundamentals of Mars' Schumann resonance, and lasting for similar minute-long timescales.  However, while \citet{Ruf:2009uy} saw evidence of non-thermal emission in the form of increased levels of kurtosis across all eight of their frequency bands between 8470 and 8490 MHz during periods of activity in the power spectrum of the kurtosis, we saw no such increase in either the spectral kurtosis or accumulated power at 8.0 GHz.  Additionally, \citet{Ruf:2009uy} reports only one instance of non-thermal emission and its corresponding signal in the kurtosis power spectrum that occurred during a large-scale dust storm; in our own Mars observations, we observe this signal in the kurtstrum many times despite there being no large-scale storms on Mars during our observations (though some dust activity and minor storms were present).

In contrast to 8.0 GHz, the accumulated power and spectral kurtosis at 3.2 GHz do show increased levels over a narrow range of frequencies, evidence that the signals in the kurtstrum are caused by narrowband RFI (Figure \ref{total}).  Non-Mars observations were done on other astrophysical sources (during testing of the instrument and for short periods prior to the Mars observations -- see Table \ref{obstable}), during which no such signals were seen in the kurtstrum.  However, our lack of off-source data during observations of Mars and the comparatively short amount of time spent on non-Mars observations means we cannot definitively say whether or not the source of the narrowband RFI is localized to Mars or terrestrial in origin. During the approximately 4.5 hours of overlap observations on 16 April 2010 between our own observations and the DSN observations, the latter saw no significant increase in kurtosis levels (Kuiper, private communication).  The multiple instances of interference in our own data at this epoch indicate that the RFI we observed during this period was most likely local.

\section{Conclusion} \label{conclusion}

Between 9 March and 2 June 2010, 30 hours of simultaneous observations of Mars at 3.2 and 8.0 GHz were conducted with the Allen Telescope Array to search for signals indicative of electrostatic discharge on Mars, following a report by \citet{Ruf:2009uy} of the detection of non-thermal emission at 8.5 GHz, coinciding with a 35 km deep dust storm.  A special wideband signal processor was developed for these observations using the tools and infrastructure of the CASPER group at the University of California, Berkeley.  The spectrometer was tested on a number of astrophysical and terrestrial sources, including PSR B0531+21 and a radar satellite.  The 1024-channel spectrometer calculates the accumulated power and accumulated power squared in real time as a function of frequency with a 104.8 MHz bandwidth; the spectral kurtosis was computed post-observation.  We calculated the power spectrum of the kurtosis, or kurtstrum, to determine if low-frequency modulation of the kurtosis occurred during Martian dust storm events and whether it corresponded with predicted frequency ranges of Mars' Schumann resonances.

We did not detect any non-thermal emission associated with electrostatic discharge.  There were limited amounts of dust activity present in the form of small dust storms, however there were no large-scale dust storms.  Signals in the kurtstrum with spectral peaks near 10 Hz and higher harmonics did occur frequently during our Mars observations (on the order of one to two events per hour of observation) in both frequency bands centered at 3.2 and 8.0 GHz.  However, these signals were shown to correspond to narrowband RFI that showed up strongly in the accumulated power and spectral kurtosis at 3.2 GHz.

In addition to showing that narrowband RFI can produce strong periodic signals in the kurtstrum over a wide range of frequencies, we have discussed the use of the spectral kurtosis as a means for detecting astrophysical phenomenon, in addition to RFI excision.  We show that kurtosis can be a powerful tool for the detection of strongly variable sources such as giant pulses from the Crab pulsar.

\acknowledgements

The authors would like to thank Christopher Ruf, Nilton O. Renno, and Thomas Kuiper for their suggestions and discussion regarding the analysis and results in this paper.  The authors would also like to acknowledge the generous support of the Paul
 G. Allen Family
 Foundation, who have provided major support for design, construction,
 and operations of
 the ATA. Contributions from Nathan Myhrvold, Xilinx Corporation, Sun
 Microsystems,
 and other private donors have been instrumental in supporting the ATA.
 The ATA has been
 supported by contributions from the US Naval Observatory in addition
 to National Science
 Foundation grants AST-050690, AST-0838268, and AST-0909245.  Support for this project was provided by Director's Research and Development Fund (DRDF) 1332 from the Jet Propulsion Laboratory (JPL).

\clearpage

\bibliography{references}

\appendix

\clearpage

\noindent
\begin{table}[h!]
\begin{center}
\begin{tabular*}{.99\textwidth}{| c | c | c | c | c |}
  \hline \hline
  Date & Time [UTC] & Source & Obs Freq [MHz] & Martian Weather \\
  & & & [BF1, BF2] & \\ \hline \hline
  \multirow{2}{*}{09 March 2010} & 01:00--01:30 & PSR B0329+54 & 3200, 8000 & \multirow{2}{4cm}{\small{Increased dust storm activity along seasonal north polar cap edge.}} \\
  & 01:30--05:30 & Mars & 3200, 8000 & \\ 
  & & & & \\ \hline
  \multirow{2}{*}{23 March 2010} & 00:30--00:45 & PSR B0329+54 & 1420, 1420 & \multirow{2}{4cm}{\small{A few minor, short-lived dust storms.}} \\
  & 01:00--05:30 & Mars & 3200, 8000 & \\ \hline
  \multirow{2}{*}{30 March 2010} & 00:20--00:30 & PSR B0329+54 & 1420, 1420 & \multirow{2}{4cm}{\small{Little dust activity; local dust storms near southern mid-latitudes.}} \\
  & 01:15--05:30 & Mars & 3200, 8000 & \\
  & & & & \\ \hline
  \multirow{3}{*}{12--13 April 2010} & 22:40--22:50 & PSR B0329+54 & 1420, 1420 & \multirow{3}{4cm}{\small{Dust activity in the north; local dust storms in the south.}} \\
  & 23:15--23:25 & W3 & 1660, 6660 & \\
  & 23:45--05:15 & Mars & 3200, 8000 & \\ \hline
  \multirow{3}{*}{16--17 April 2010} & 22:40--22:50 & PSR B0329+54 & 1420, 1420 & \multirow{3}{4cm}{\small{Increased dust activity in the north; local dust storms in the south.}} \\
  & 23:10--23:20 & W3 & 1660, 6660 & \\
  & 23:45--05:20 & Mars & 3200, 8000 & \\ \hline
  \multirow{3}{*}{08--09 May 2010} & 22:05--22:15 & PSR B0329+54 & 1420, 1420  & \multirow{3}{4cm}{\small{Little dust activity; local dust clouds in southern mid-latitudes.}} \\
  & 22:35--22:50 & W3 & 1660, 6660 & \\
  & 23:15--04:45 & Mars & 3200, 8000 & \\ \hline
  \multirow{3}{*}{01--02 June 2010} & 22:20--22:25 & PSR B0329+54 & 1420, 1420 & \multirow{3}{4cm}{\small{Little to no dust activity.}} \\
  & 22:45--22:50 & W3 & 1660, 6660 & \\
  & 23:15--03:20 & Mars & 3200, 8000 & \\ \hline
  \hline
\end{tabular*}
\caption{Table of observation dates and sources.}
\tablecomments{BF1 and BF2 refer to beamformers 1 and 2.  W3 refers to observations of masers in the Westerhout-3 star forming region.  All weather information taken from NASA/JPL-Caltech/Malin Space Science Systems \citep{website:malin1, website:malin2, website:malin3, website:malin4, website:malin5, website:malin6}.\label{obstable}}
\end{center}
\end{table}

\noindent
\begin{table}[h!]
\begin{center}
\begin{tabular*}{.895\textwidth}{| c | c | c | c | c |}
  \hline \hline
  & \multicolumn{2}{|c|}{Antenna Count} & \multicolumn{2}{|c|}{Synthesized SEFD [Jy]} \\ \hline
  \multirow{2}{*}{Epoch} & 3.2 GHz & 8.0 GHz & 3.2 GHz & 8.0 GHz \\
  & [X-pol, Y-pol] & [X-pol, Y-pol] & [X-pol, Y-pol] & [X-pol, Y-pol] \\ \hline \hline
  09 March 2010 & 26, 24 & 26, 25 & 502.31, 639.16 & 874.09, 732.87 \\ \hline
  23 March 2010 & 24, 24 & 25, 24 & 542.81, 639.16 & 920.00, 769.44 \\ \hline
  30 March 2010 & 25, 24 & 25, 24 & 529.09, 639.16 & 920.00, 769.44 \\ \hline
  12--13 April 2010 & 23, 18 & 22, 17 & 567.37, 841.73 & 1047.19, 1070.93 \\ \hline
  16--17 April 2010 & 23, 18 & 24, 21 & 567.37, 841.73 & 970.29, 885.65 \\ \hline
  08--09 May 2010 & 22, 18 & 23, 19 & 603.18, 872.74 & 967.62, 951.32 \\ \hline
  01--02 June 2010 & 24, 17 & 25, 19 & 525.57, 814.55 & 875.82, 948.23 \\ \hline
  \hline
\end{tabular*}
\caption{Effective system equivalent flux density (SEFD) for both polarizations and observing frequencies for each observation date. \label{senstable}}
\end{center}
\end{table}

\noindent
\begin{table}[h!]
\begin{center}
\begin{tabular*}{.902\textwidth}{| c | c | c | c | c | c |}
  \hline \hline
  Integration & Frequency & Channel & Total & Total Mars & Beam Width \\ 
  Time & Channels & Bandwidth & Bandwidth & Obs Time & [3.2 GHz, 8.0 GHz] \\ \hline \hline
  1.25 ms & 1024 & 0.1 MHz & 104.8 MHz & 30 hours & 100\arcsec, 40\arcsec \\
  \hline
\end{tabular*}
\caption{High time resolution kurtosis spectrometer (HiTREKS) parameters. \label{tab:hitreks_specs}}
\end{center}
\end{table}

\end{document}